\documentclass[aps,pra,twocolumn,superscriptaddress,floatfix,10pt, preprintnumbers]{revtex4-2}

\usepackage{amsmath,amssymb,amsfonts,mathtools,bm,graphicx}
\usepackage[colorlinks=true,linkcolor=blue,citecolor=magenta]{hyperref}
\usepackage{orcidlink}

\newcommand{\Sp}{\hat S^{+}}
\newcommand{\Sm}{\hat S^{-}}
\newcommand{\Sz}{\hat S^{z}}
\newcommand{\ad}{\hat a^{\dagger}}
\newcommand{\X}{\hat X}
\newcommand{\nm}{\hat n^{\mathrm{m}}}
\newcommand{\Deff}{\Delta_{\mathrm{eff}}}
\newcommand{\Vb}{V_{\mathrm{bind}}}
\newcommand{\half}{\tfrac12}
\newcommand{\Imm}{\operatorname{Im}}
\newcommand{\Ree}{\operatorname{Re}}
\newcommand{\Tr}{\operatorname{Tr}}

\providecommand{\ket}[1]{\left|#1\right\rangle}
\providecommand{\bra}[1]{\left\langle#1\right|}

\begin{document}

\title{Dressed magnon dynamics in a Bose-Hubbard bath:\\ retardation, pairing, entanglement and two-magnon scattering}

\author{Andrés N. Cáliz\,\orcidlink{0009-0002-6938-8595}}
\email{andres.navas@qilimanjaro.tech}
\affiliation{Qilimanjaro Quantum Tech, Carrer de Vene\c{c}uela 74, 08019 Barcelona, Spain}
\affiliation{Departament de Física, Universitat de Barcelona, 08007 Barcelona, Spain}

\author{Arnau Riera\,\orcidlink{0000-0002-3271-7802}}
\affiliation{Qilimanjaro Quantum Tech, Carrer de Vene\c{c}uela 74, 08019 Barcelona, Spain}

\author{Enrique Rico\,\orcidlink{0000-0003-4414-6821}}
\affiliation{Theoretical Physics Department, CERN, 1211 Geneva 23, Switzerland}

\author{Jo\~{a}o Barata\,\orcidlink{0000-0003-4286-4555}}
\affiliation{Theoretical Physics Department, CERN, 1211 Geneva 23, Switzerland}

\author{Marcin P\l odzie\'n\,\orcidlink{0000-0002-0835-1644}}
\email{marcin.plodzien@qilimanjaro.tech}
\affiliation{Qilimanjaro Quantum Tech, Carrer de Vene\c{c}uela 74, 08019 Barcelona, Spain}

\date{\today}

\preprint{CERN-TH-2026-191}

\begin{abstract}
We study one- and two-magnon dynamics in a spin-$1/2$ XX chain coupled to a tunable Bose-Hubbard bath. The XX Hamiltonian transports the magnons, while the bath dresses their motion and mediates an effective attraction. A single magnon forms a mobile polaron whose retarded bosonic cloud reduces its velocity below the static Lang-Firsov prediction and generates magnon-boson entanglement. In the weak-dressing, off-resonant mobile regime, the residual velocity deficit and entanglement are governed by the excitation weight of the cloud, as captured by perturbation theory and tensor network-based real-time simulations. Transport therefore provides a model-calibrated proxy for entanglement without joint-state reconstruction. When two clouds overlap, the bath mediates a finite-range attraction, yielding finite-size evidence for a compact two-magnon bound state and a transient post-scattering signature. On-site bath interactions suppress local boson number fluctuations and stiffen the bath response, reducing entanglement and weakening ground-state binding. Bath deformability thus emerges as a common control parameter for magnon transport, entanglement, and bath-mediated pairing.

\end{abstract}

\maketitle

\section{Introduction}\label{sec:intro}

When a mobile excitation propagates through a many-body medium, such as an energetic electron passing through an ion cloud, a new quasiparticle state, i.e., a polaron~\cite{Landau,Feynman62}, can be formed by the dressing of the excitation by the modes of the bulk. While non-dynamical properties of polarons, such as band narrowing, effective mass, and the phase diagram of Holstein bipolarons~\cite{Alexandrov,Franchini2021,Sous2018,Pistolesi2021}, are well understood, the real-time description of these states remains incomplete. To this end, quantum platforms such as Bose and Fermi polarons in ultracold gases~\cite{Zwerlein2009,Jorgensen2016,Camargo2018}, together with excitation-phonon coupling engineered in trapped ions and Rydberg-atom arrays~\cite{Solano2012,Cirac2012,Plodzien2017,Plodzien2018,Kosior2023}, might allow us to study the real-time properties of polarons and the dynamical formation of these quasiparticles, including exotic phenomena such as phonon-assisted string breaking~\cite{Mallick2025} and the flow of quantum resources between spins, bosons, and their correlations~\cite{Plodzien2026Sensitivity}.

A medium with a finite response time cannot follow a moving excitation instantly. Capturing this delay effect is, in general, challenging because eliminating the bath generates a temporally non-local interaction, making the dynamics dependent on the entire history of the propagating state. Indeed, the lag reduces the quasiparticle's velocity below its band-narrowed value and, at the same time, adds dynamical correlations to the entanglement already generated by the position-dependent static dressing. The former effect is a transport observable, while the latter is a property of the joint quantum state, ordinarily out of reach without reconstructing the full bath. That the residual slowdown and the entanglement are both controlled by the dressing excitation weight, once the static contribution is accounted for, suggests that the measurable velocity deficit may provide an experimentally accessible proxy for excitation-bath entanglement.

In this work, we test this observation in a minimal spin-boson model: a spin-$1/2$ XX chain carrying one or two magnons, each coupled to a Bose-Hubbard bath through a Holstein term. In the absence of the bath, a single magnon propagates freely and two hard-core magnons scatter without binding, so every effect discussed below (dressing, retardation, entanglement, and dynamical attraction between magnons) must originate from the interaction with the bath. Unlike typical solid-state environments~\cite{Holstein1959b,Holstein1959,Alexandrov}, an engineered bath allows the on-site repulsion to be tuned. Increasing this repulsion suppresses local boson-number fluctuations and stiffens the bath response~\cite{SchmidtKoch2013,Hartmann2016}, thereby controlling its deformability. We analyze the system using both a Davydov coherent-state mean field~\cite{Davydov1969,Davydov1977,Davydov1973b,Davydov1976}, valid in a classical picture, and converged tensor network (matrix product state [MPS]) simulations to account for correlations beyond the restricted single-field Davydov manifold.

In the case of a single magnon, the retardation and the entanglement rise and saturate together, both fixed by the number of excitations in the dressing cloud, so that the measurable velocity deficit tracks the entanglement up to a slowly varying logarithmic correction. For two magnons, the dressing clouds overlap, lowering the bath energy when each cloud reaches the other magnon. This XXZ-like attraction, absent for a single magnon, produces finite-size ground-state evidence for a small bound molecule and a transient compact component during a scattering process. The same on-site repulsion suppresses the dressing cloud, causing both the entanglement and the binding to decrease. Their parallel behavior reflects their common dependence on the bath’s deformability.

This paper is organized as follows. Section~\ref{sec:model} sets up the model and its polaron frame. In Section~\ref{sec:single}, we treat the single-magnon state, from the mean-field equations to the retardation-entanglement relation. We extend this discussion to the two-magnon case in Section~\ref{sec:bind}, where we study the attractive potential in equilibrium and during scattering processes. We conclude in Section~\ref{sec:disc} with a summary and discussion of the main results. Details of the analytical derivations and numerical results are collected in the appendices.

\section{Magnons in a spin-$1/2$ XX chain coupled to a Bose-Hubbard bath}\label{sec:model}
We consider a lattice system made of a one-dimensional spin-$1/2$ XX chain carrying one or two magnons, coupled through a Holstein term~\cite{Holstein1959,Plodzien2026Sensitivity} to a bath of the Bose-Hubbard type~\cite{Fisher1989}. The Hamiltonian of this system can be written as $\hat H=\hat H_{\mathrm{s}}+\hat H_{\mathrm{B}}+\hat H_{\mathrm{int}}$, with
\begin{subequations}\label{eq:H}
\begin{align}
  \hat H_{\mathrm{s}}&=-\frac J2\sum_j\big(\Sp_j\Sm_{j+1}+\mathrm{H.c.}\big),\\
  \hat H_{\mathrm{B}}&=-t_{\mathrm{B}}\sum_j\big(\ad_j\hat a_{j+1}+\mathrm{H.c.}\big)
   +\omega_0\sum_j\hat n_j\nonumber\\
   &\quad+\frac U2\sum_j\hat n_j(\hat n_j-1),\\
  \hat H_{\mathrm{int}}&=\lambda_z\sum_j\big(\ad_j+\hat a_j\big)\nm_j \, ,
\end{align}
\end{subequations}
where $\nm_j=\half-\Sz_j$, $\hat n_j=\ad_j\hat a_j$, $\hbar=1$, and $\hat H_{\mathrm{s}}$ ($\hat H_{\mathrm{B}}$) is the spin-chain (bath) Hamiltonian. The bath couples to the magnon density $\nm_j$ through the interaction Hamiltonian $\hat H_{\mathrm{int}}$, and $\nm_j=0$ on the polarized background and $\nm_j=1$ on a flipped site, so the magnon-free chain exerts no force on the bath. Coupling to $\Sz_j$ instead would differ only by a uniform bath displacement, immaterial at $U=0$ but not at $U>0$. We take the density coupling because it leaves the magnon-free background insensitive to any bath-induced force. Unless otherwise stated, the mobile magnon calculations use $J$ as the energy unit, $J^{-1}$ as the time unit, and unit lattice spacing.

Since $[\hat H,\sum_j\Sz_j]=0$, the total magnetization is conserved and we can work in sectors with fixed magnon number $\sum_j\nm_j=1,2$, with a single bare magnon dispersing as $\varepsilon(k)=-J\cos k$. The bath is a tight-binding lattice of bosons with band $\omega(q)=\omega_0-2t_{\mathrm{B}}\cos q$, which is gapped and stable for $t_{\mathrm{B}}<\omega_0/2$, and an on-site Bose-Hubbard interaction $U$. The Holstein coupling $\hat H_{\mathrm{int}}$ does not conserve the boson number, so the bosons are genuine bath excitations rather than conserved particles, which restricts the realizations to platforms whose bath quanta are themselves excitations; see Sec.~\ref{sec:disc}.

To separate the polaronic dressing from the residual bath-mediated interactions, we pass to the polaron frame through a Lang-Firsov transformation~\cite{LangFirsov1963,Mahan}, with $ \; \;\hat U=e^{-\hat S}$, $\hat S=\kappa\sum_j(\ad_j-\hat a_j)\nm_j$, and $\kappa=-\lambda_z/\omega_0$. This removes the linear coupling, dresses the exchange with the bond displacement operators $\X_j=e^{\kappa(\ad_j-\hat a_j)}$, and recasts the bath operators in terms of the density-shifted boson $\hat b_j=\hat U\hat a_j\hat U^\dagger=\hat a_j+\kappa\nm_j$. The exact transformed Hamiltonian (see Appendix~\ref{app:LF}) reads, up to a constant,
\begin{equation}\label{eq:HLF}
  \hat H_{\mathrm{LF}}=\hat H_J+\hat H_0+\hat H_t+\hat H_U
   -\frac{\lambda_z^2}{\omega_0}\sum_j\nm_j \, .
\end{equation}
The dressed exchange, the on-site bath energy, the bath hopping, and the bath interaction read, respectively,
\begin{subequations}\label{eq:HLFterms}
\begin{align}
  \hat H_J&=-\frac J2\sum_j\big(\Sp_j\Sm_{j+1}\,\X_j\X^\dagger_{j+1}+\mathrm{H.c.}\big),\\
  \hat H_0&=\omega_0\sum_j\ad_j\hat a_j,\\
  \hat H_t&=-t_{\mathrm{B}}\sum_j\big(\hat b^\dagger_j\hat b_{j+1}+\mathrm{H.c.}\big),\\
  \hat H_U&=\frac U2\sum_j\hat b^\dagger_j\hat b_j\big(\hat b^\dagger_j\hat b_j-1\big).
\end{align}
\end{subequations}
The final term in Eq.~\eqref{eq:HLF} is the polaron self-energy $-\lambda_z^2/\omega_0$ per magnon, which is a constant in each fixed-magnon-number sector. Only the exchange term acquires explicit displacement operators. The Hamiltonian $\hat H_J$ contains the bond operators $\X_j$ responsible for the band narrowing, while the on-site energy $\hat H_0$ remains expressed in terms of the undisplaced bosons $\hat a_j$, and the bath hopping $\hat H_t$ and interaction $\hat H_U$ retain their functional form in the shifted bosons $\hat b_j$. Expanding the shifted quadratic hopping (an exact expansion that terminates at order $\kappa^2$) gives the nearest-neighbor contribution
\begin{equation}\label{eq:Vbind}
  -\Vb\sum_j\nm_j\nm_{j+1},\qquad \Vb=2\kappa^2t_{\mathrm{B}}\, ,
\end{equation}
a nearest-neighbor magnon attraction of depth $\Vb$. In the spin language, it generates the operator $\nm_j\nm_{j+1}=\Sz_j\Sz_{j+1}$ up to single-site and constant terms. To fix the XXZ sign convention, after the staggered rotation $\hat S_j^\pm\to(-1)^j\hat S_j^\pm$, the effective spin Hamiltonian reads (up to constant terms)
\begin{equation}
  \hat H_{\mathrm{XXZ}}^{\mathrm{eff}}
  =
  \widetilde J\sum_j
  \left(
    \hat S_j^x\hat S_{j+1}^x
    +
    \hat S_j^y\hat S_{j+1}^y
    +
    \Deff\Sz_j\Sz_{j+1}
  \right) \, ,
\end{equation}
with $\widetilde J=Je^{-\kappa^2}$ and $\Deff=-\Vb/\widetilde J=-2\kappa^2t_{\mathrm{B}}/(Je^{-\kappa^2})$. Thus, $\Deff<0$ denotes the induced ferromagnetic longitudinal anisotropy in this convention. For dispersionless bosons, the polaron cloud is strictly on-site. Since two hard-core magnons never share a site, their clouds never overlap, and no interaction results.

This transformation effectively dresses each magnon with the lattice deformation it produces, of amplitude set by $\langle\nm_j\rangle$. The exchange is reduced by the overlap $e^{-\kappa^2}$ of the clouds displaced about the two ends of a bond, and acquires a Peierls phase when the bath is in motion. The induced interaction in Eq.~\eqref{eq:Vbind} originates in the same cloud, which, for a dispersive bath, spreads to neighboring sites, where it overlaps with a second magnon.
\section{Single-magnon dressing and retardation}
\label{sec:single}

A single magnon provides the simplest setting in which one can separate three effects of the bath: the static Lang-Firsov band narrowing, the additional slowdown caused by the finite response time of the dressing cloud, and the magnon-boson entanglement generated during polaron formation. We first derive a coherent-state description of the coupled dynamics and then compare it with the full many-body evolution. Unless stated otherwise, the MPS real-time protocol starts from a bare magnon wavepacket and the laboratory-frame bath vacuum. The single-field Davydov calculation uses the same spin wavepacket and $\beta_j(0)=0$ in the polaron frame; this natural variational reference is not an exact representation of the laboratory-frame vacuum for a delocalized magnon, as detailed in Appendices~\ref{app:dav} and~\ref{app:num}.

\subsection{Semiclassical dynamics}
\label{sec:davydov}

The mean-field equations can be obtained from the time-dependent variational principle
\begin{equation}
\label{eq:tdvp}
  \delta\!\int\!dt\;
  \bra{\Psi(t)}
  \big(i\partial_t-\hat H_{\mathrm{LF}}\big)
  \ket{\Psi(t)}
  =0\, ,
\end{equation}
applied to the single-field Davydov $D_2$ coherent-state ansatz~\cite{Davydov1969,Zhao1997,Kosior2023,Plodzien2018},
\begin{equation}
\label{eq:ansatz1}
  \ket{\Psi(t)}
  =
  \Big(\sum_j\psi_j(t)\,\Sm_j\Big)\ket{\Uparrow}
  \otimes
  \prod_j
  e^{\beta_j(t)\ad_j-\beta_j^*(t)\hat a_j}
  \ket{0}.
\end{equation}
Here, $\ket{\Uparrow}$ is the fully polarized state,
$\rho_j=|\psi_j|^2$ is the magnon density with $\sum_j\rho_j=1$, and $\beta_j$ is the coherent bath amplitude in the polaron frame. The corresponding laboratory-frame amplitude is
\begin{equation}
  \gamma_j
  \equiv
  \langle\hat a_j\rangle_{\mathrm{lab}}
  =
  \beta_j+\kappa\rho_j,
  \qquad
  \kappa=-\frac{\lambda_z}{\omega_0}.
\end{equation}
Thus, $\kappa$ measures the displacement produced by an occupied spin site, while the dynamical adiabaticity is controlled by the bath frequencies relative to the dressed magnon bandwidth.

Stationarity with respect to the complex canonical variables
$(\psi_j,\psi_j^*)$ and $(\beta_j,\beta_j^*)$ gives
\begin{subequations}
\label{eq:eom1}
\begin{align}
  i\dot\psi_j
  ={}&
  -\frac J2 e^{-\kappa^2}
  e^{i(\theta_j-\theta_{j+1})}\psi_{j+1}
  \nonumber\\
  &
  -\frac J2 e^{-\kappa^2}
  e^{i(\theta_j-\theta_{j-1})}\psi_{j-1}
  +\Phi_j\psi_j,
  \label{eq:eom1a}\\
  i\dot\beta_j
  ={}&
  \omega_0\beta_j
  -t_{\mathrm{B}}(\gamma_{j-1}+\gamma_{j+1})
  \nonumber\\
  &
  +i\kappa\big(
  \mathcal J_{j,j+1}-\mathcal J_{j-1,j}
  \big)
  \nonumber\\
  &
  +U\rho_j|\beta_j+\kappa|^2(\beta_j+\kappa)
  +U(1-\rho_j)|\beta_j|^2\beta_j,
  \label{eq:eom1b}
\end{align}
\end{subequations}
where
\begin{align}
  \theta_j
  &=
  2\kappa\Imm\beta_j,\\
  \Phi_j
  &=
  -2t_{\mathrm{B}}\kappa\,\Ree(\beta_{j-1}+\beta_{j+1})
  \nonumber\\
  &\quad
  +\frac U2
  \big(
  |\beta_j+\kappa|^4-|\beta_j|^4
  \big),
  \label{eq:Vj}\\
  \mathcal J_{j,j+1}
  &=
  Je^{-\kappa^2}\Imm\!\big[
  e^{i(\theta_j-\theta_{j+1})}
  \psi_j^*\psi_{j+1}
  \big].
  \label{eq:current}
\end{align}
The first equation is a tight-binding Schr\"odinger equation with the Lang-Firsov reduction
$J\mapsto Je^{-\kappa^2}$, a bath-induced Peierls phase, and the potential $\Phi_j$.
The second is a driven discrete Gross-Pitaevskii equation. Its current-divergence source
shows explicitly how magnon motion excites the bath, while $\Phi_j$ and $\theta_j$ describe
the reciprocal action of the bath on the magnon. The $U$ term is the exact coherent-state
average over the two local configurations contained in the ansatz: an occupied site with
field $\beta_j+\kappa$ and an empty site with field $\beta_j$; see Appendix~\ref{app:dav}.

It is useful to distinguish the coherent contribution $|\gamma_j|^2$ to the bath occupation from the total bath
occupation predicted by the same ansatz. In the laboratory frame,
\begin{equation}
\label{eq:d2occupation}
  \langle\hat n_j\rangle_{D_2}
  =
  |\gamma_j|^2
  +
  \kappa^2\rho_j(1-\rho_j).
\end{equation}
The second term is the local fluctuation associated with superposing different magnon
positions. We use $|\gamma_j|^2$ below to visualize the classical deformation, but comparisons
with the many-body occupation must use Eq.~\eqref{eq:d2occupation}. The single-field ansatz is
expected to be most accurate for a fast bath and weak dressing. The condition
$\max_j|\gamma_j|^2<1$ is therefore used only as a practical diagnostic.

\subsection{Dressing, retardation, and the mobile regime}
\label{sec:retard}

Let $\Omega_{\mathrm{B}}=\omega_0-2t_{\mathrm{B}}$ denote the bottom of the noninteracting bath band. When
$\Omega_{\mathrm{B}}$ is large compared with the dressed magnon scale
$\widetilde J=Je^{-\kappa^2}$, the bath follows the magnon quasi-instantaneously and
$\dot\beta_j$ may be neglected in Eq.~\eqref{eq:eom1b}. For $U=0$ and a wavepacket centered near the bottom of the dressed band and broad
compared with the bath-response length, eliminating the bath produces the focusing nonlinear
Schr\"odinger equation \mbox{(see Appendix~\ref{app:nls})}:
\begin{equation}
\label{eq:nls}
  i\partial_t\psi
  =
  -\frac{1}{2m^*}\partial_x^2\psi
  -g|\psi|^2\psi,
  \qquad
  g=
  \frac{8\kappa^2t_{\mathrm{B}}^2}{\omega_0-2t_{\mathrm{B}}}\, ,
\end{equation}
where the lattice spacing is set to unity, as is done for continuum expressions. This particular self-focusing mechanism requires a dispersive bath: it vanishes for the Einstein bath ($t_{\mathrm{B}}=0$) used in the following to isolate polaron formation. In the dispersive case, we restrict the simulations to the mobile side of the self-trapping crossover, where the nonlinearity changes the packet width without localizing its center of mass. The Lang-Firsov transformation gives the reference group velocity
\begin{equation}
\label{eq:v1}
  v_{\mathrm{LF}}(k_0)
  =
  Je^{-\kappa^2}\sin k_0\, ,
\end{equation}
corresponding to the velocity of an instantaneously dressed magnon. When the bath response is not instantaneous, the deformation trails the packet and the measured velocity falls below $v_{\mathrm{LF}}(k_0)$. The relevant softness is set by $\Omega_{\mathrm{B}}$ and by the displacement $|\kappa|$, rather than by $\omega_0$ alone.

\begin{figure}[t]
  \centering
  \includegraphics[
    width=\columnwidth
  ]{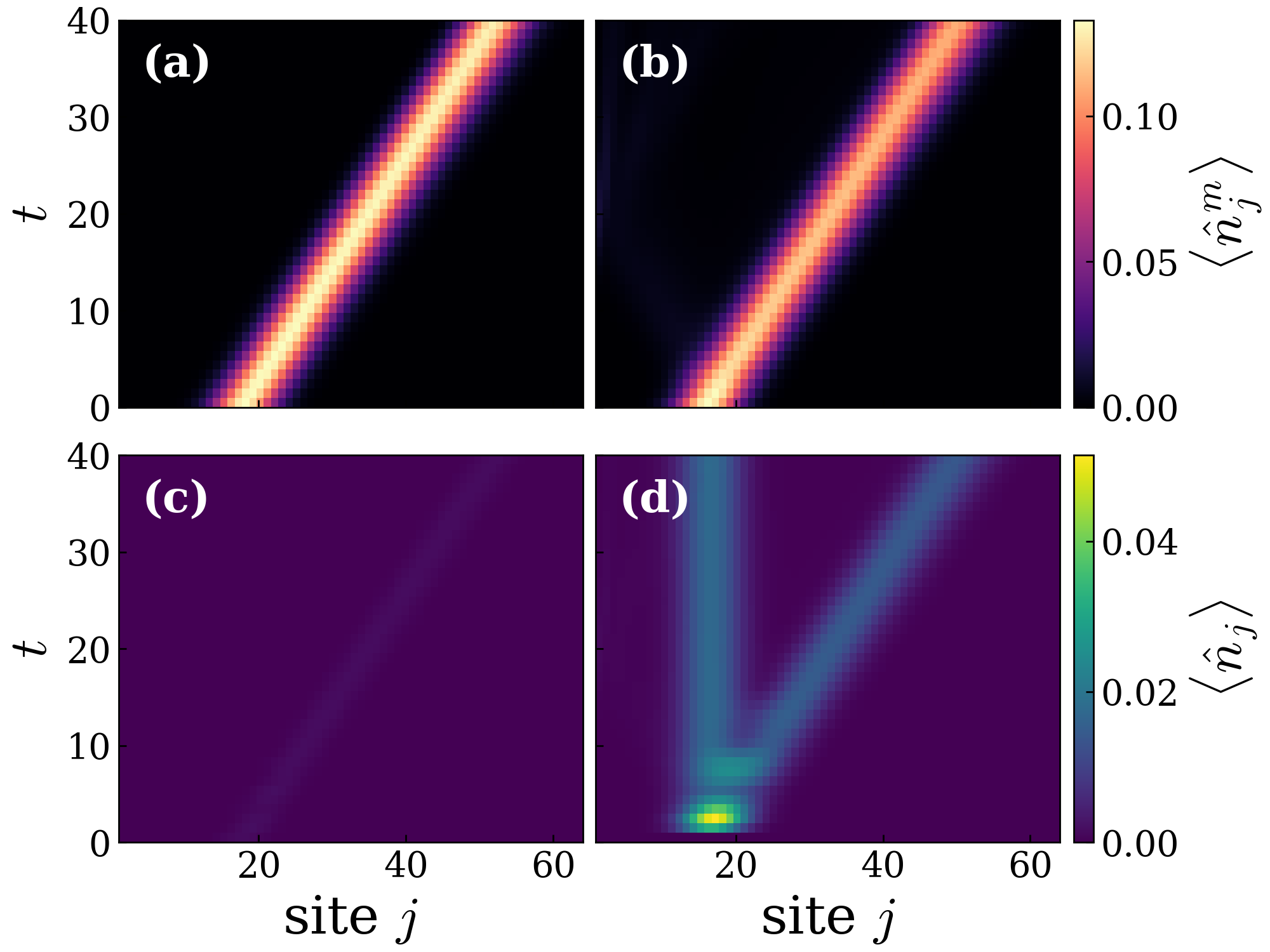}
\caption{Spatial anatomy of single-magnon retardation: comparison of the Davydov mean field with the
many-body dynamics at matched parameters ($\omega_0=2$, $\lambda_z=0.6$, $t_{\mathrm{B}}=0$, $U=0$, $k_0=\pi/2$,
$L=64$). (a,b) Magnon density $\langle\nm_j\rangle$ in the mean field (a) and the many-body dynamics (b). The two results are
nearly indistinguishable: the mean field captures the dressed transport, while the small residual
slowdown reflects the finite bath response time. (c,d) Bath occupation from the
coherent cloud $|\gamma_j|^2$ (c) and the many-body $\langle \hat n_j\rangle$ (d), using a common color scale. The
mean-field cloud is substantially fainter than the many-body one: the coherent field carries
only the mean-field contribution and misses the quench transient at the release site (oscillating at
$\omega_0$ as the cloud forms) and the stationary occupation shed when the packet departs. This
excess in the many-body occupation, absent from (c), signals dynamics beyond the coherent contribution. The dispersionless bath ($t_{\mathrm{B}}=0$) is chosen here for clarity, while a dispersive bath adds the effects discussed in Sec.~\ref{sec:bind}.}
  \label{fig:single}
\end{figure}

In Fig.~\ref{fig:single}, we compare this coherent-state picture with the many-body dynamics for an Einstein bath, with both calculations essentially reproducing the same ballistic magnon packet.
The bath sector is more revealing: the coherent contribution $|\gamma_j|^2$ follows the moving packet,
whereas the many-body occupation also contains a release-site transient and bath excitations
left behind by the departing polaron. These features are absent from $|\gamma_j|^2$ and signal dynamics beyond the coherent contribution. However, quantitatively assessing
the mean-field error in bath occupation requires a comparison with the full
Davydov prediction in Eq.~\eqref{eq:d2occupation}, rather than with $|\gamma_j|^2$ alone.

\subsection{Magnon-boson entanglement}

We quantify the entanglement between the spin chain and the bath using the von Neumann entropy
\begin{align}
\label{eq:Smbdef}
  S_{\mathrm{mb}}(t)
  &=
  -\Tr\!\left[
  \rho_{\mathrm{s}}(t)\ln\rho_{\mathrm{s}}(t)
  \right],
  \nonumber\\
  \rho_{\mathrm{s}}(t)
  &=
  \Tr_{\mathrm{b}}
  \ket{\Psi(t)}\bra{\Psi(t)}.
\end{align}
Because the full spin-boson state remains pure, $S_{\mathrm{mb}}$ is the entanglement entropy
across the magnon-boson bipartition. A magnon pinned at a definite site displaces a definite
bath mode and need not become entangled with it. By contrast, a delocalized magnon correlates
each position with a different displaced bath state. Their overlap, which is smaller than unity, suppresses the
off-diagonal elements of $\rho_{\mathrm{s}}$ and generates entanglement; see Appendix~\ref{app:collapse}.

\begin{figure}[t]
  \centering
  \includegraphics[
    width=0.9\columnwidth
  ]{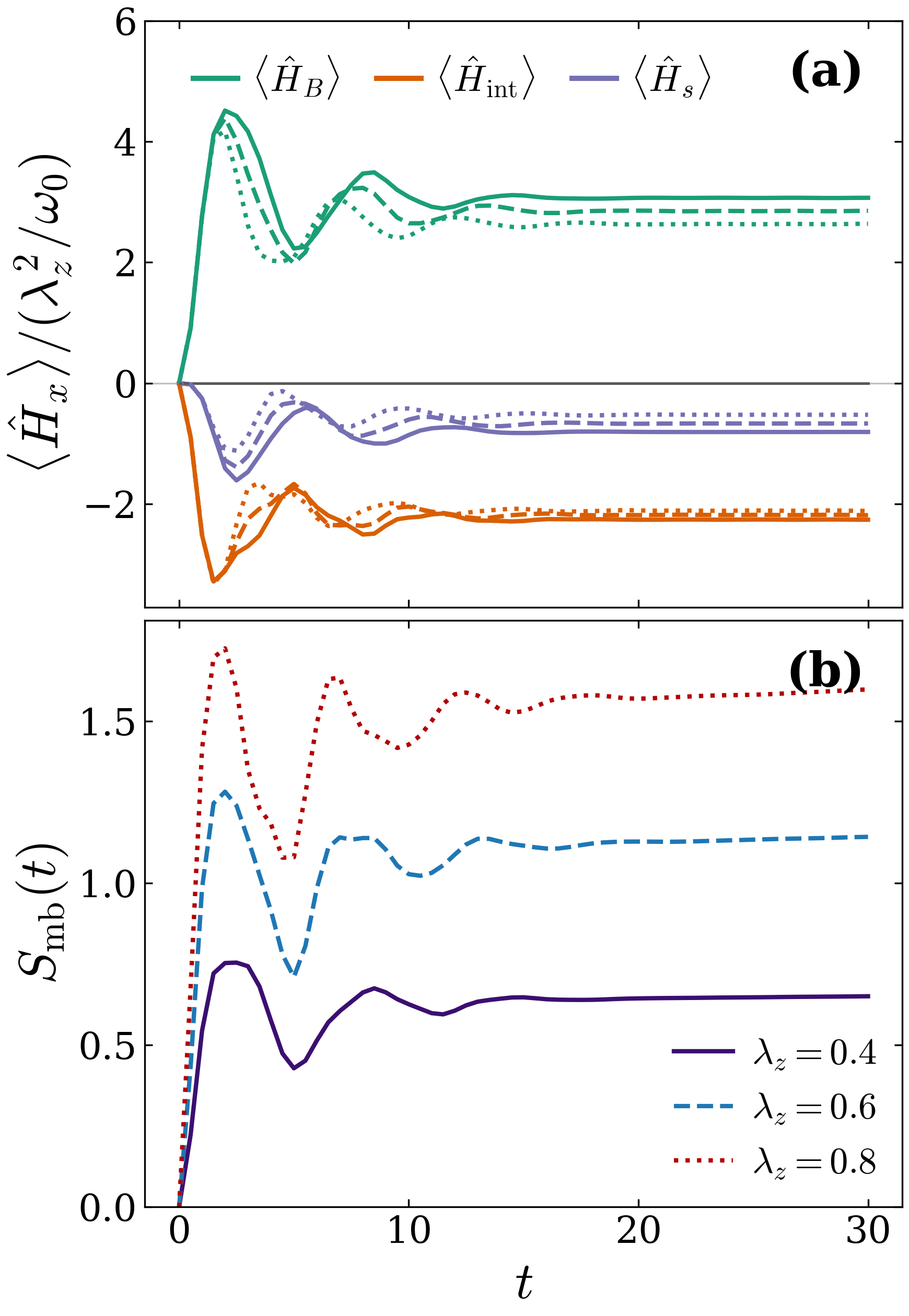}
  \caption{Formation of a single-magnon polaron after release with the bath initially in its vacuum
  ($\omega_0=2$, $k_0=\pi/2$, $L=64$) for
  $\lambda_z=0.4, 0.6, 0.8$ (solid, dashed, dotted).
  (a) Bath, interaction, and exchange energies in units of
  $\lambda_z^2/\omega_0$, together with the conserved total energy (gray).
  (b) Magnon-boson entanglement entropy for the same couplings.
  }
  \label{fig:dynamics}
\end{figure}

Starting from the bare product state, $S_{\mathrm{mb}}(t)$ rises while the cloud forms, overshoots
as the bath rings at $\omega_0$, and then fluctuates around a plateau; see Fig.~\ref{fig:dynamics}(b). Since a finite closed system does not possess a strict $t\rightarrow\infty$ limit, we denote by $S_{\mathrm{mb}}^*$ the average over the late-time window specified in Appendix~\ref{app:num}. The same formation process is visible in the energy budget in Fig.~\ref{fig:dynamics}(a). For a packet centered at $k_0=\pi/2$, the initial spin
exchange energy vanishes up to finite-width corrections, while the vacuum gives $\langle\hat H_{\mathrm{B}}\rangle=\langle\hat H_{\mathrm{int}}\rangle=0$. During dressing, $\langle\hat H_{\mathrm{B}}\rangle$ becomes positive, the interaction energy becomes negative, and the
exchange sector readjusts so that the total energy remains constant. The oscillations in all three sectors and in $S_{\mathrm{mb}}(t)$ occur on the same bath timescale.

\subsection{Velocity-entanglement relation}

The common origin of the slowdown and the entanglement becomes explicit at weak coupling. For an Einstein bath and $\omega_0>J$, a bare plane-wave magnon of momentum $k_0$ couples, to first order in $\lambda_z$, only to states containing one virtual bath excitation,
\begin{align}
\label{eq:pert_main}
  \ket{\Psi(t)}
  &=
  c_0(t)\ket{k_0;0}
  +
  \sum_q c_q(t)\ket{k_0-q;1_q},
\end{align}
with $\overline{|c_q|^2}= \frac{2\lambda_z^2}{L\Delta_q^2}$ and $\Delta_q=\varepsilon_{k_0} -\varepsilon_{k_0-q}- \omega_0$, where the overline denotes an average over the post-transient time window. The total one-boson weight is $\bar n=\sum_q\overline{|c_q|^2}$. Replacing the oscillatory probabilities by their late-time averages gives the dephased
perturbative entropy
\begin{equation}
\label{eq:Smb_main}
  S_{\mathrm{mb}}^{\mathrm{PT}}
  =
  -(1-\bar n)\ln(1-\bar n)
  -
  \sum_q
  \overline{|c_q|^2}
  \ln\overline{|c_q|^2}.
\end{equation}
This expression is exact for the dephased one-boson density matrix at this order. For the finite wavepacket obtained numerically, we evaluate this expression directly using the
same packet and averaging window rather than an infinite plane wave.

\begin{figure}[t]
  \centering
  \includegraphics[
    width=\columnwidth
  ]{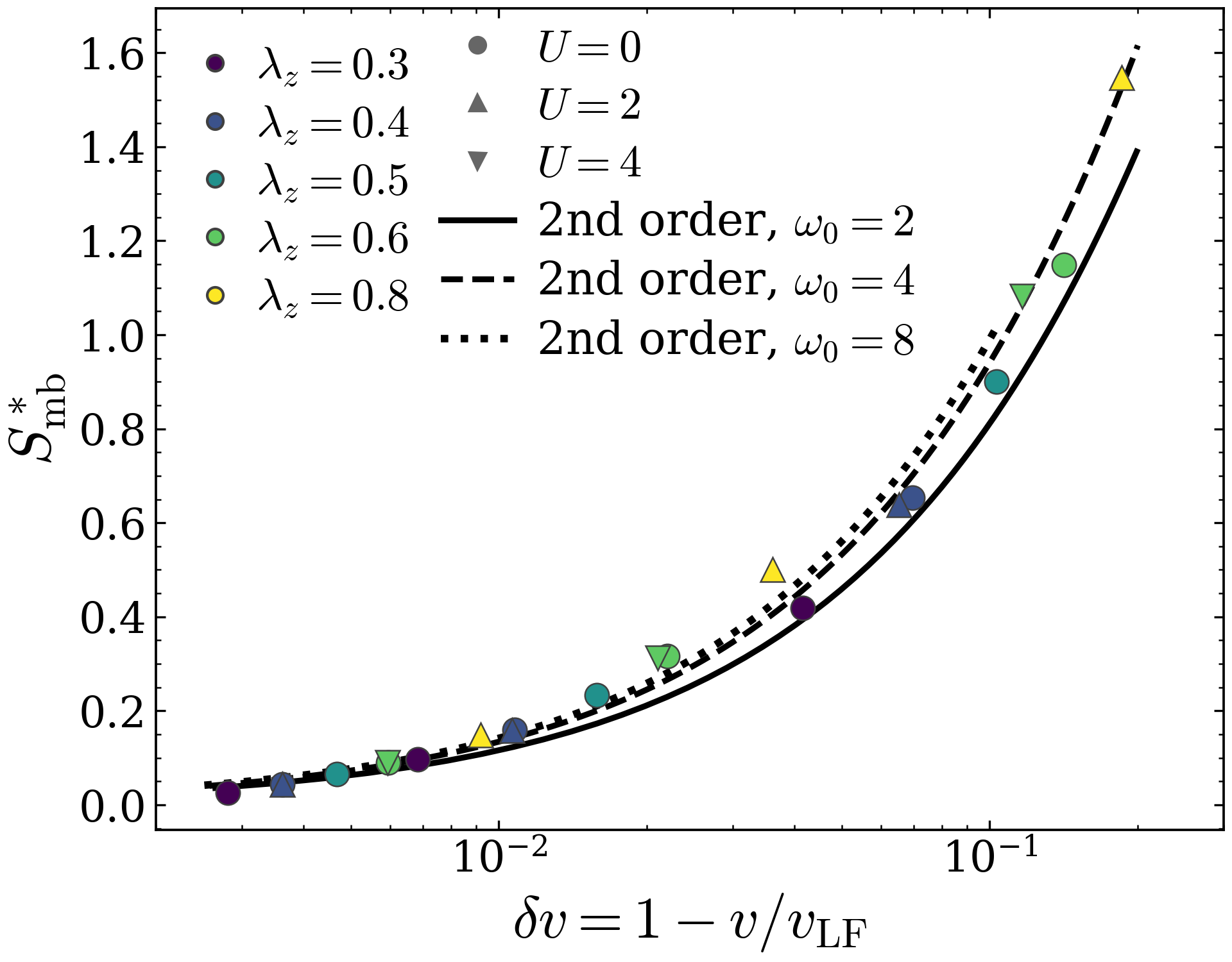}
  \caption{
  Late-time magnon-boson entanglement versus the velocity deficit for a single
  magnon ($k_0=\pi/2$, $L=64$). Colors denote $\lambda_z$ and markers denote $U$.
  The deficit is corrected with a $\lambda_z=0$ packet of identical width; see Appendix~\ref{app:num}. Dashed curves show second-order perturbation theory results for
  $\omega_0=2,4,8$. The numerical data include the same three bath frequencies, which are not distinguished by additional marker styles. The curves are evaluated for the simulated wavepacket without fitted parameters.
  The collapse holds in the weak-dressing, off-resonant regime and deteriorates near $\omega_0=J$ and near self-trapping.
  }
  \label{fig:retard}
\end{figure}

At $k_0=\pi/2$, the same weights determine the magnon recoil. Defining the plane-wave velocity deficit relative to Eq.~\eqref{eq:v1} as $\delta v=1-\bar v/v_{\mathrm{LF}}$, perturbation theory gives
\begin{equation}
\label{eq:collapse_main}
  \delta v
  =
  \bar n-\kappa^2+O(\kappa^4),
  \qquad
  \bar n
  =
  \frac{2\kappa^2}
  {\big[1-(J/\omega_0)^2\big]^{3/2}}.
\end{equation}
For a finite packet, the numerical definition includes the $\lambda_z=0$ control factor given in Appendix~\ref{app:num}. Both $\bar n$ and $\kappa^2$ are of order $\lambda_z^2$. Therefore, the subtraction in Eq.~\eqref{eq:collapse_main} must be retained even in the antiadiabatic limit. Writing $\overline{|c_q|^2}=\bar n p_q$, with $\sum_q p_q=1$, yields
\begin{align}
\label{eq:Sstruct_main}
  S_{\mathrm{mb}}^{\mathrm{PT}}
  &=
  \bar n\big[
  \mathcal A-\ln\bar n
  \big]
  +
  O(\bar n^2),
  \nonumber\\
  \mathcal A
  &=
  1+H_p,
  \qquad
  H_p
  =
  -\sum_q p_q\ln p_q.
\end{align}
The logarithm corresponds to the entropy cost of distributing a small total dressing weight among the resolvable bath modes. For a finite wavepacket, its momentum resolution regularizes the spurious $\ln L$ dependence of the plane-wave expression. Over the off-resonant parameter range of Fig.~\ref{fig:retard}, the normalized shape $p_q$ changes more slowly than $\bar n$. Consequently, both the deficit and the entropy are governed primarily by the same dressing weight, producing the approximate collapse in Fig.~\ref{fig:retard}. The relation is not universal: it retains a weak dependence on the packet shape and on $\omega_0$ through $H_p$, and it fails near resonance, where real bath excitations can be emitted.

The same perturbative counting explains the weak dependence on $U$ for a single magnon. The on-site interaction acts only on states with at least two bath excitations. Such states enter observables at order $\lambda_z^4$, so $U$ does not modify the leading one-boson relation. Its effect becomes appreciable for the shared, multiboson cloud of the two-magnon bound state, which we discuss in Section~\ref{sec:bind}.

Thus, within the mobile and off-resonant regime, the measured slowdown provides an experimentally accessible proxy for the otherwise nonlocal magnon-boson entanglement. The proxy is quantitative only after specifying the packet and the bath parameters. Outside this regime, particularly when on-shell emission or self-trapping becomes possible, velocity and entanglement need not follow the same relation.
\section{Two-magnon bath-mediated interaction and binding}
\label{sec:bind}

Two hard-core magnons on the bare chain scatter without binding~\cite{Bethe1931,Wortis1963}. Consequently, any bound state found in the present model originates from the bosonic bath, in contrast to the exchange-bound magnon pairs realized in optical lattices~\cite{Fukuhara2013}. Each magnon locally deforms a dispersive bath. When the two deformations overlap, the bath energy is lowered, producing an effective attraction between the magnons. This interaction has no single-magnon counterpart and, in the appropriate limit, can be expressed as an effective XXZ
anisotropy. We first characterize the induced potential and the resulting equilibrium bound state, and then examine how the same mechanism modifies the properties of real-time scattering processes.

\subsection{Born-Oppenheimer approximation}
\label{sec:Vr}

To isolate the bath-mediated interaction, we use the Born-Oppenheimer construction described in Appendix~\ref{app:BO}. We set $J=0$, thereby freezing the magnons and treating them as static sources $\lambda_z(\ad_j+\hat a_j)$, and then relax the bath around each source configuration. The construction is exact in the heavy-magnon limit, in which every $\nm_j$ is conserved, and provides a controlled description when the bath is fast relative to the magnon dynamics, $\omega_0-2t_{\mathrm{B}}\gg J$. The dependence of the relaxed bath ground-state energy on the separation between the two sources defines the effective potential $V(r)$.

\subsubsection{Bath-mediated potential}

The nearest-neighbor attraction in Eq.~\eqref{eq:Vbind} is the leading contribution to a longer-range interaction. For noninteracting bosons ($U=0$), the bath Hamiltonian is quadratic. For an $L$-site periodic bath, displacing its normal modes gives the ground-state energy exactly, and the component that depends on the magnon separation is
\begin{equation}
\label{eq:Vr}
\begin{split}
  V(r)
  &=
  -2\lambda_z^2\,G_r,
  \qquad
  z=
  \frac{
  \omega_0-\sqrt{\omega_0^2-4t_{\mathrm{B}}^2}
  }{2t_{\mathrm{B}}},
  \\
  G_r
  &=
  \frac{1}{L}
  \sum_q
  \frac{\cos(qr)}
  {\omega_0-2t_{\mathrm{B}}\cos q}
  \xrightarrow{L\to\infty}
  \frac{
  z^{|r|}
  }{
  \sqrt{\omega_0^2-4t_{\mathrm{B}}^2}
  }.
\end{split}
\end{equation}
Thus, in the thermodynamic limit, the potential is proportional to the square of the coupling and to the static bath susceptibility $G_r$ and decays exponentially as $V(r)=A\,e^{-r/r_0}$, with
\begin{equation}
\label{eq:Ar0}
  A=
  \frac{
  -2\lambda_z^2
  }{
  \sqrt{\omega_0^2-4t_{\mathrm{B}}^2}
  },
  \qquad
  r_0=
  \frac{1}{\ln(1/z)}.
\end{equation}
At $r=1$ in the antiadiabatic regime
($t_{\mathrm{B}}\ll\omega_0$, $G_1\to t_{\mathrm{B}}/\omega_0^2$), Eq.~\eqref{eq:Vr} reduces to $V(1)\to-2\kappa^2t_{\mathrm{B}}=-\Vb$, in agreement with Eq.~\eqref{eq:Vbind}. The full expression extends
this nearest-neighbor result to all separations and throughout the stable regime $t_{\mathrm{B}}<\omega_0/2$. At fixed bath dispersion, $\lambda_z$ controls the amplitude $A\propto\lambda_z^2$ without changing the decay length. By contrast, changing $t_{\mathrm{B}}/\omega_0$ modifies the bath susceptibility and hence affects both the range and the magnitude of the response. The range increases as the bath softens, diverges as $t_{\mathrm{B}}\to\omega_0/2$, and vanishes in the Einstein-bath limit ($t_{\mathrm{B}}\to0$, $r_0\to0$), where the response becomes strictly local and no interaction is induced between hard-core magnons on distinct sites.

\begin{figure}[t]
  \centering
  \includegraphics[
    width=0.9\columnwidth
  ]{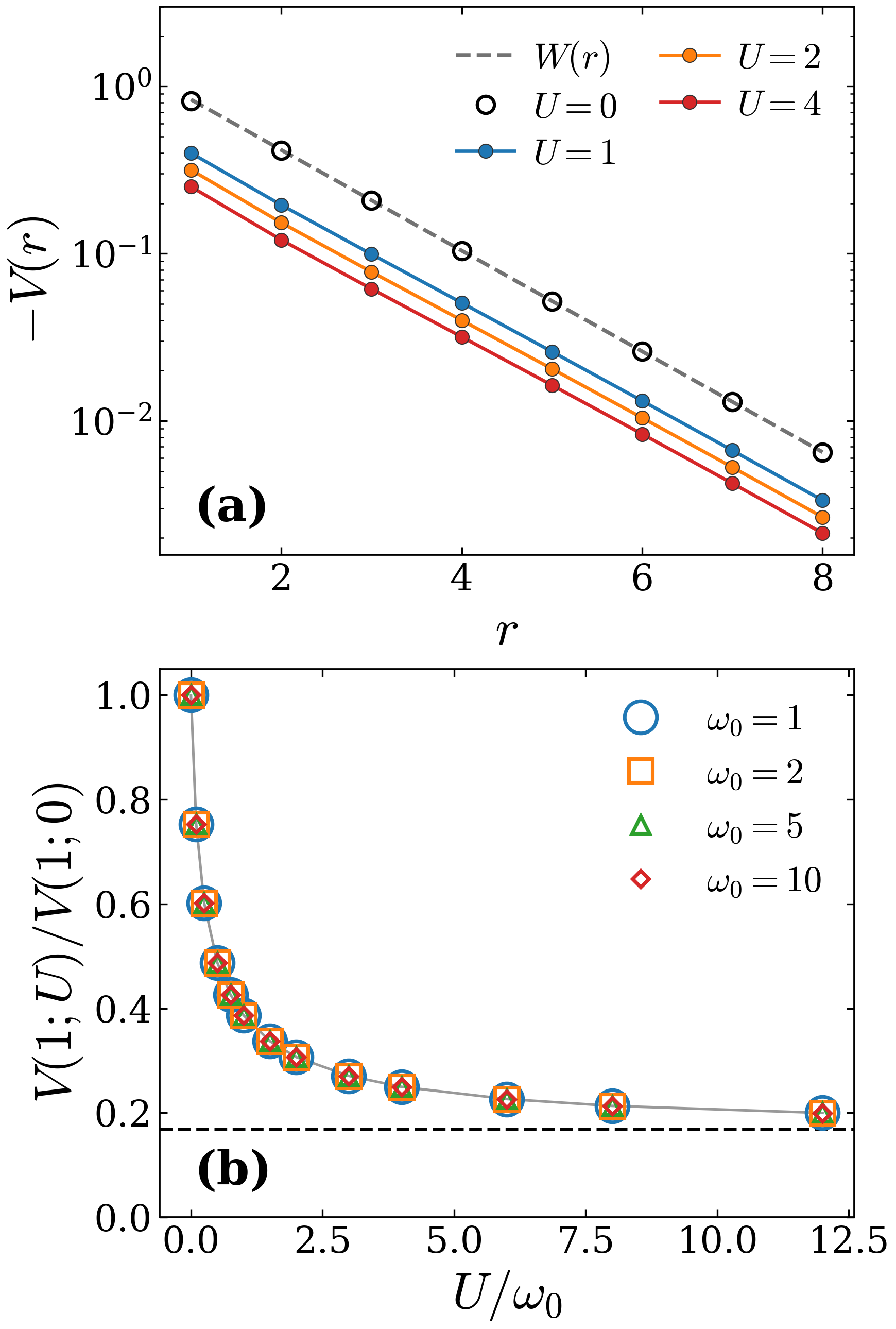}
  \caption{
  Bath-mediated potential in the Born-Oppenheimer limit ($J=0$), with
  $\lambda_z/\omega_0=0.5$ and $t_{\mathrm{B}}/\omega_0=0.4$.
  (a) Logarithmic plot of $-V(r)$ at $\omega_0=2$ for $U=0$-$4$.
  The open circles at $U=0$ reproduce the analytic result $W(r)$
  (dashed line); increasing $U$ lowers the approximately parallel profiles.
  (b) Depth ratio $V(1;U)/V(1;0)$ as a function of $U/\omega_0$
  for $\omega_0=1$-$10$ (nested open markers). The ratio approaches
  the finite hard-core limit $f_\infty$ of Eq.~\eqref{eq:floor}
  (black dashed line), with $f_\infty\simeq0.17$ for the parameters shown.
  }
  \label{fig:VU}
\end{figure}

Equation~\eqref{eq:Vr} is exact only for noninteracting bosons. When the Bose-Hubbard interaction $U$ is nonzero, the bath is anharmonic, and we evaluate the potential numerically using the same Born-Oppenheimer construction. Density matrix renormalization group (DMRG)~\cite{White1992} yields the bath ground-state energy for three static configurations:
no sources, one source, and two sources separated by $r$. Their mixed second difference,
\begin{equation}
\label{eq:Vsub_num}
  V(r;U)
  =
  E(\{0,r\})
  -2E(\{0\})
  +E(\varnothing),
\end{equation}
removes the bath vacuum energy and the individual dressing energy of each magnon, thereby isolating their interaction. Equivalently, this calculation corresponds to the full spin-boson
chain at $J=0$ with the two magnons pinned. For $U=0$, the numerical result reproduces Eq.~\eqref{eq:Vr} within the DMRG truncation error. The linear profiles on the logarithmic scale in
Fig.~\ref{fig:VU}(a) confirm the exponential decay, with the range controlled by $t_{\mathrm{B}}/\omega_0$ and the amplitude controlled by $\lambda_z$. For $U>0$, where the harmonic closed-form expression no longer applies, the same subtraction yields the anharmonic potential directly.

Within the explored parameter range, the principal effect of $U$ is an approximately uniform rescaling of the potential. Increasing $U$ reduces the depth of the attractive well, while the
log-linear profiles remain approximately parallel, motivating the representation
\begin{equation}
\label{eq:VfU}
  V(r;U)
  \simeq
  f(U)\,W(r),
  \qquad
  W(r)\equiv V(r;0),
\end{equation}
where $f(0)=1$ and the fitted factor is approximately independent of $r$; see Fig.~\ref{fig:VU}(a). The shorthand $f(U)$ is used only at fixed $t_{\mathrm{B}}/\omega_0$ and $\lambda_z/\omega_0$, as it is not a universal function of $U$ alone. In the Born-Oppenheimer limit, $\omega_0$ sets the bath energy scale. At fixed $t_{\mathrm{B}}/\omega_0$ and $\lambda_z/\omega_0$, the suppression is therefore organized by the dimensionless ratio $U/\omega_0$, as indicated by the collapse of the curves for $\omega_0=1$-$10$ in Fig.~\ref{fig:VU}(b). As $U$ increases, $f(U)$ decreases from unity and
approaches a finite floor rather than vanishing,
\begin{equation}
\label{eq:floor}
  f_\infty
  \equiv
  \lim_{U\to\infty}f(U)
  =
  \frac{
  V(1;\infty)
  }{
  V(1;0)
  }.
\end{equation}
The free-fermion mapping of the hard-core bath, presented in Appendix~\ref{app:Uinf}, accounts for this nonzero limit. At $\lambda_z/\omega_0=0.5$, we obtain $f_\infty\simeq0.17$, shown by the dashed line in Fig.~\ref{fig:VU}(b). Thus, the bath interaction substantially weakens the induced attraction but does not fully eliminate it.

The saturation can be understood analytically in the hard-core limit $U\to\infty$; see Appendix~\ref{app:Uinf}. In this limit, the bath bosons map to free fermions whose single-particle dispersion, $\omega(q)=\omega_0-2t_{\mathrm{B}}\cos q$, is the same as that of the harmonic modes. Consequently, the leading-order response has the same spatial decay governed by the lattice Green's function $G_r$. The hard-core constraint limits the local occupation response and reduces the potential depth, while the numerical profiles retain essentially the same range. The depth therefore approaches a finite fraction of its $U=0$ value as $U\to\infty$. This behavior differs from the limit $t_{\mathrm{B}}\to0$, which removes the bath dispersion and hence the interaction between distinct sites.

\subsubsection{Bound-state condition}

In the antiadiabatic limit, $\omega_0-2t_{\mathrm{B}}\gg J$, the bath responds on a timescale much shorter than that of the magnons and can be integrated out. The resulting effective model is a static
ferromagnetic XXZ chain. Its hopping is narrowed to $\widetilde J=Je^{-\kappa^2}$, while the nearest-neighbor attraction $\Vb$ produces the nominal $U=0$ anisotropy
\begin{equation}
\label{eq:Deff}
  \Deff
  =
  -\frac{\Vb}{\widetilde J}
  =
  -\frac{
  2\kappa^2t_{\mathrm{B}}
  }{
  Je^{-\kappa^2}
  }.
\end{equation}
For $U>0$, the corresponding nearest-neighbor estimate is instead
$\Deff(U)\simeq f(U)\Deff$. The corresponding two-magnon problem at $U=0$ is exactly solvable using the coordinate Bethe ansatz~\cite{Bethe1931,Wortis1963}. At fixed total momentum $K$, the center-of-mass coordinate separates according to $\psi_{jk}=e^{iK(j+k)/2}\phi(r)$, with the bound-state solution corresponding to an exponentially localized relative wavefunction. The relative coordinate then experiences a single attractive well of depth $\Vb$ and a $K$-dependent hopping $\tau=\widetilde J\cos(K/2)$. As shown in Appendix~\ref{app:two}, a bound state exists when the
attraction overcomes the relative kinetic scale,
\begin{equation}
\label{eq:thr}
  |\Deff|>\cos(K/2),
  \qquad
  E_b=
  \frac{
  (\Vb-\tau)^2
  }{
  \Vb
  },
\end{equation}
where $\tau=Je^{-\kappa^2}\cos(K/2)$ and $E_b$ is the energy by which the bound state lies below the two-magnon continuum. In the nearest-neighbor antiadiabatic model, the $K=0$ channel has the widest relative band ($\tau=\widetilde J$), and the bound-state condition reduces to the Ising regime $|\Deff|>1$.

At $K=0$, considered below in the scattering analysis, this effective description predicts binding when
$|\Deff|=\Vb/\widetilde J=2\kappa^2t_{\mathrm{B}}e^{\kappa^2}/J>1$, reflecting the competition between the induced attraction $\Vb=2\kappa^2t_{\mathrm{B}}$ and the narrowed hopping $\widetilde J=Je^{-\kappa^2}$. Three ingredients govern this competition. The dressing strength $|\kappa|=|\lambda_z|/\omega_0$ controls the attraction through $\Vb\propto\kappa^2$. A finite bath hopping $t_{\mathrm{B}}$ is required because an Einstein bath does not mediate interactions between distinct sites, while stability requires $t_{\mathrm{B}}<\omega_0/2$ so that the bath band remains gapped. A nonzero $U$ reduces the induced attraction through the factor $f(U)$ shown in Fig.~\ref{fig:VU}, thereby shifting the effective binding threshold. Strong dressing simultaneously narrows the magnon bandwidth and can substantially reduce the pair mobility. It is therefore useful to distinguish the existence of a bound pair from its ability to propagate: a compact pair may remain bound while becoming increasingly heavy.

\subsection{Binding in the full quantum bosonic bath}

The threshold in Eq.~\eqref{eq:thr} is a nearest-neighbor, antiadiabatic estimate. We now retain the full quantum bath and determine the binding directly, without integrating out the bosons. Because the induced two-body term vanishes in the single-magnon sector, $\sum_j\nm_j\nm_{j+1}=0$, any additional two-magnon binding is generated by the bath. Alongside
the binding energy, we use the separation distribution
$g(r)=\sum_j\langle\nm_j\nm_{j+r}\rangle$ and its nearest-neighbor component $P_{\mathrm{nn}}=g(1)$ to characterize the compactness of the pair.

\subsubsection{Binding energy in the full bath}

We compute the binding energy of the mobile pair directly from the DMRG ground-state energies in the sectors with fixed magnon number,
\begin{equation}
\label{eq:Eb_dmrg}
  E_b
  =
  2(E_1-E_0)
  -(E_2-E_0),
\end{equation}
where $E_0$, $E_1$, and $E_2$ are the ground-state energies for zero, one, and two magnons, respectively. In a finite system, $2(E_1-E_0)$ provides the reference formed from two independent single-magnon addition energies, and $E_b(L)>0$ is finite-size evidence for pair binding. A thermodynamic bound state would require the persistence of this positive value under finite-size scaling. Unlike in the Born-Oppenheimer construction, the magnon separation is not imposed, and the pair selects its own relative wavefunction, whose
spatial extent is described by $g(r)$. All ground-state data in this subsection use $L=16$.

\begin{figure}[t]
  \centering
  \includegraphics[
    width=0.9\linewidth
  ]{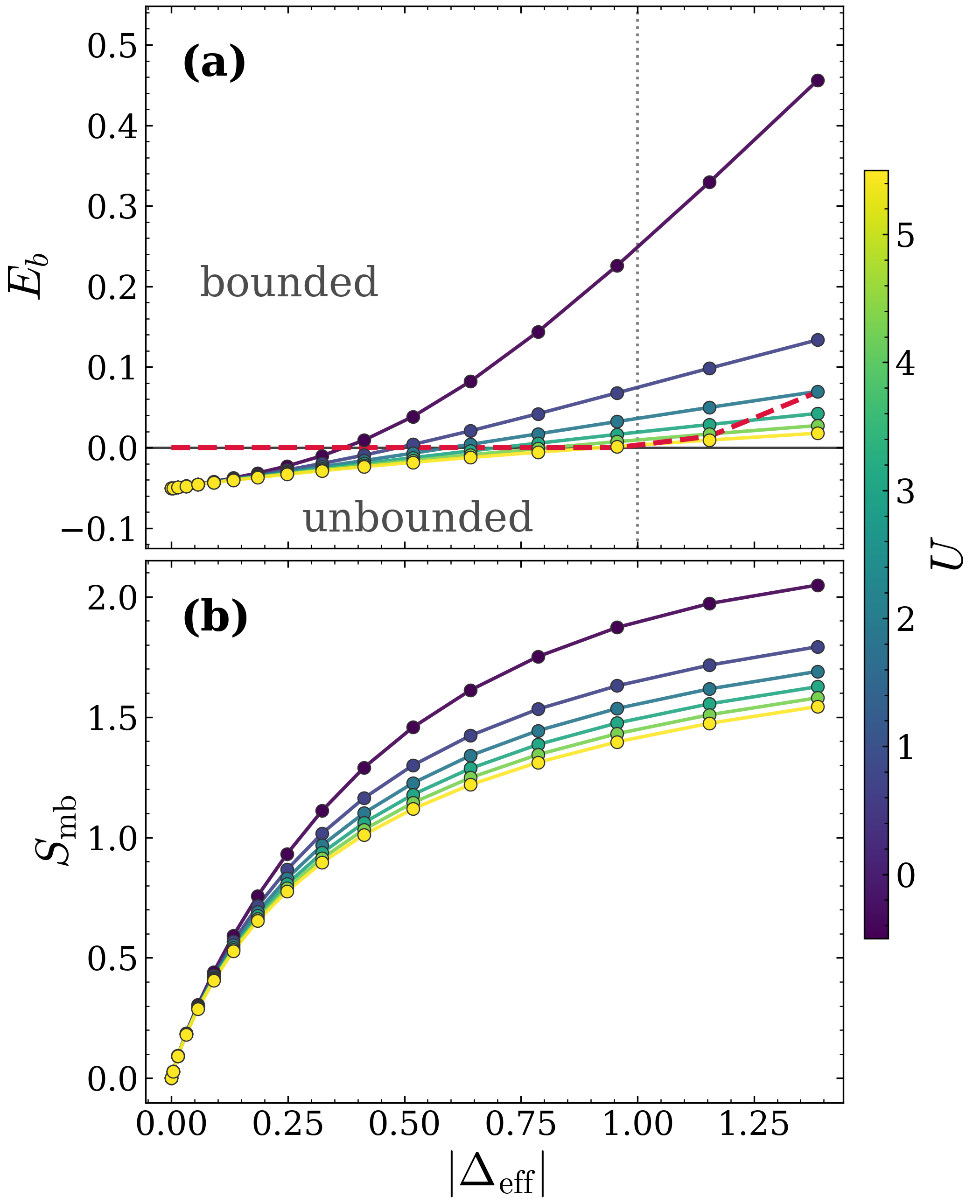}
  \caption{
  Ground-state properties of two magnons coupled to the full quantum bath
  ($\omega_0=3$, $t_{\mathrm{B}}=1$, $L=16$). The nominal $U=0$ antiadiabatic anisotropy $|\Deff|$
  is varied through $\lambda_z$, and colors denote six values of $U$.
  (a) Binding energy $E_b$ defined in Eq.~\eqref{eq:Eb_dmrg}.
  The dashed curve is the nearest-neighbor antiadiabatic estimate of
  Eq.~\eqref{eq:thr}, and the dotted vertical line marks $|\Deff|=1$.
  Increasing $U$ reduces $E_b$ and shifts its onset to larger $|\Deff|$.
  (b) Magnon-boson entanglement entropy $S_{\mathrm{mb}}$, which increases
  with dressing and is suppressed by $U$.
  }
  \label{fig:binding}
\end{figure}

Figure~\ref{fig:binding} summarizes the results for the full quantum bath. Within the range of dressings shown, a positive $E_b(L)$ provides finite-size evidence for binding. The condition
$|\Deff|=1$ serves only as an approximate reference because the actual induced potential has finite range and the parameters satisfy $\omega_0-2t_{\mathrm{B}}=J$, outside the asymptotically
antiadiabatic regime. Consistently, the full-bath calculation yields finite-size binding evidence at smaller values of $|\Deff|$ than predicted by the nearest-neighbor estimate. As the pair becomes more compact, $P_{\mathrm{nn}}$ increases from its free value to values above $0.5$. The relative wavefunction nevertheless extends over several lattice sites. Near the onset, $g(r)$ reaches its maximum at $r=2$ rather
than at nearest-neighbor separation, reflecting the finite spatial range of the attractive well (data for $P_{\mathrm{nn}}$ and $g(r)$ are not shown).

Increasing $U$ produces the opposite trend: both the binding energy and the adjacent-pair weight decrease. This behavior is consistent with the reduction of the static potential depth described by $f(U)$ in Fig.~\ref{fig:VU}. At $\lambda_z=1.5$, $E_b(L)$ remains positive for $U\lesssim2$ and becomes negative by $U\simeq3$. Thus, for the parameters explored here, the finite-size binding signature is favored by a deformable, weakly interacting bath and is suppressed as the on-site repulsion stiffens the bosonic response.

\subsubsection{Ground-state entanglement}

The natural bipartition for the composite system separates the spin chain, restricted to the two-magnon sector, from the bosonic bath, rather than attempting to partition the two indistinguishable magnons. Because the full
ground state is pure, the von Neumann entropy of the reduced spin state, $S_{\mathrm{mb}}=-\Tr[\rho_{\mathrm{s}}\ln\rho_{\mathrm{s}}]$, with $\rho_{\mathrm{s}}=\Tr_{\mathrm{b}}\ket{\Psi_0}\bra{\Psi_0}$, measures the magnon-boson entanglement without ambiguity. As shown in Fig.~\ref{fig:binding}(b), it increases from zero
at $\lambda_z=0$ to approximately $2$ at $\lambda_z=2$, while increasing $U$ suppresses it. The entropy therefore quantifies the bosonic dressing of the magnon configurations and vanishes for a product state across the magnon-boson partition.

This entropy is not, by itself, a diagnostic of pair binding: two spatially separated dressed magnons may also be entangled with the bath. Its dependence on $\lambda_z$ and $U$ instead reflects
the deformability of the bath. A dressed spin configuration is correlated with a superposition of local boson-number states whose characteristic width scales as $\Delta n\sim|\gamma|$. The on-site term $\tfrac{U}{2}\hat n(\hat n-1)$ penalizes large number fluctuations, squeezing the local response
toward a narrower occupation distribution and reducing $S_{\mathrm{mb}}$. At $U=0$, the bath is most deformable and the entropy is largest. In the hard-core limit ($U\to\infty$; see Appendix~\ref{app:Uinf}), the local occupation is restricted, but the bath response remains finite. Accordingly, $S_{\mathrm{mb}}$ is expected to approach a reduced but nonzero value rather than
vanishing. The parallel suppression of $S_{\mathrm{mb}}$ and of the binding energy is therefore consistent with their common dependence on the bath response, although the two quantities
measure distinct physical properties.

\subsection{Magnon-magnon scattering in the interacting bosonic bath}

\begin{figure}[h]
  \centering
  \includegraphics[
    width=\columnwidth
  ]{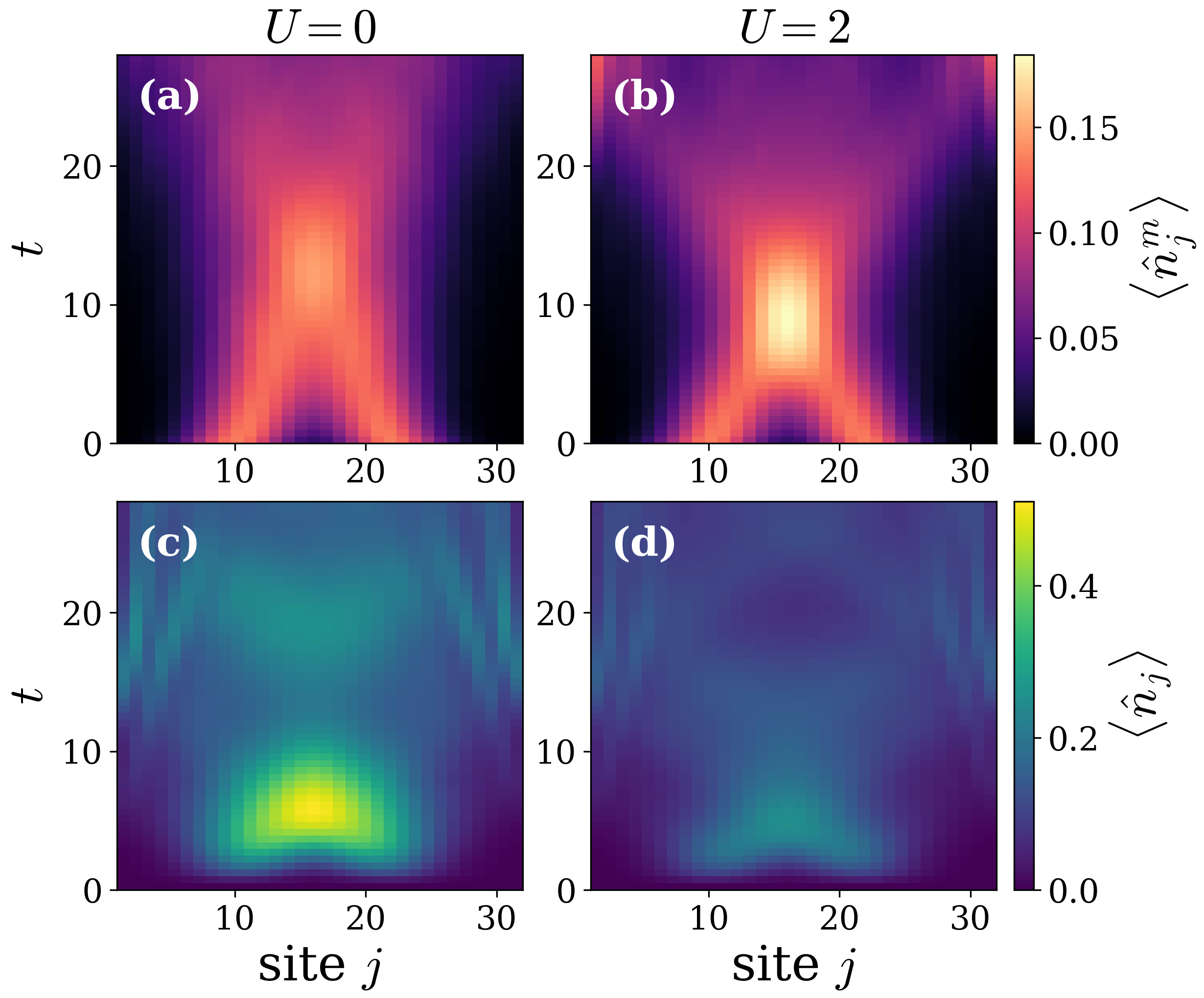}
  \caption{
  Two-magnon collision at $K=0$ for $U=0$ (left) and $U=2$ (right), with
  $\lambda_z=1.2$, $t_{\mathrm{B}}=0.8$, $\omega_0=2$, $k_0=\pi/2$, $L=32$, and
  $\chi_{\max}=128$.
  Top: magnon density $\langle\nm_j\rangle$.
  Bottom: bath occupation $\langle\hat n_j\rangle$.
  The wavepackets meet near $t\simeq7$. The magnon density retains a
  compact post-collision component in both cases, whereas the bath occupation
  is substantially reduced for $U=2$. Corresponding panels use common color
  scales, and the displayed interval extends to $t=28$.
  }
  \label{fig:collision}
\end{figure}

Having characterized the equilibrium bound state, we now examine how the same bath-mediated attraction modifies the scattering between two initially separated magnon wavepackets. Real-time scattering of bound states has been extensively explored in recent years in related models~\cite{SuraceLerose2021,Karpov2022,Milsted2022,Su2024,Turco2024,
Belyansky2024,Papaefstathiou2025,Jha2025,Farrell2025,Davoudi2025,
BarataQian2026}. Here, the introduction of the bath qualitatively distinguishes our study from these previous works, which do not include such a bulk medium; see, e.g.,~\cite{BarataRico2026,BarataQian2023} for a related discussion. The packets
are launched with opposite momenta, so that their total momentum is $K=0$, and are evolved using matrix product states (MPS), as shown in Fig.~\ref{fig:collision}. In the absence of coupling to
the bath, the two density wavepackets pass through one another and subsequently separate. Coupling to a dispersive bath changes this behavior: part of the colliding amplitude remains spatially compact and
propagates with a shared bosonic deformation. We refer to this component as the captured weight.

For the parameters displayed in Fig.~\ref{fig:collision},
$\lambda_z=1.2$ and $|\Deff|=0.83$. The nearest-neighbor antiadiabatic estimate in Eq.~\eqref{eq:thr} therefore lies below its $K=0$ threshold. Because that criterion neglects the
finite range and retardation of the full bath, it does not by itself determine whether the post-collision component corresponds to a stationary or transient state of the complete Hamiltonian. We
therefore use the neutral term \textit{captured weight} and restrict the interpretation to the simulated time window.

Figure~\ref{fig:collision} compares the scenarios $U=0$ and $U=2$. In both cases, the magnon density remains concentrated after the wavepackets meet. The bath occupation, however, is substantially smaller for $U=2$ than for $U=0$. This comparison indicates that the on-site interaction suppresses the bosonic response while preserving a compact post-collision magnon component over the displayed time interval. The trend is consistent with the ground-state results in Fig.~\ref{fig:binding}, where increasing $U$ reduces both the binding energy and the magnon-boson entanglement.

\begin{figure}[t]
  \centering
  \includegraphics[
    width=0.9\columnwidth
  ]{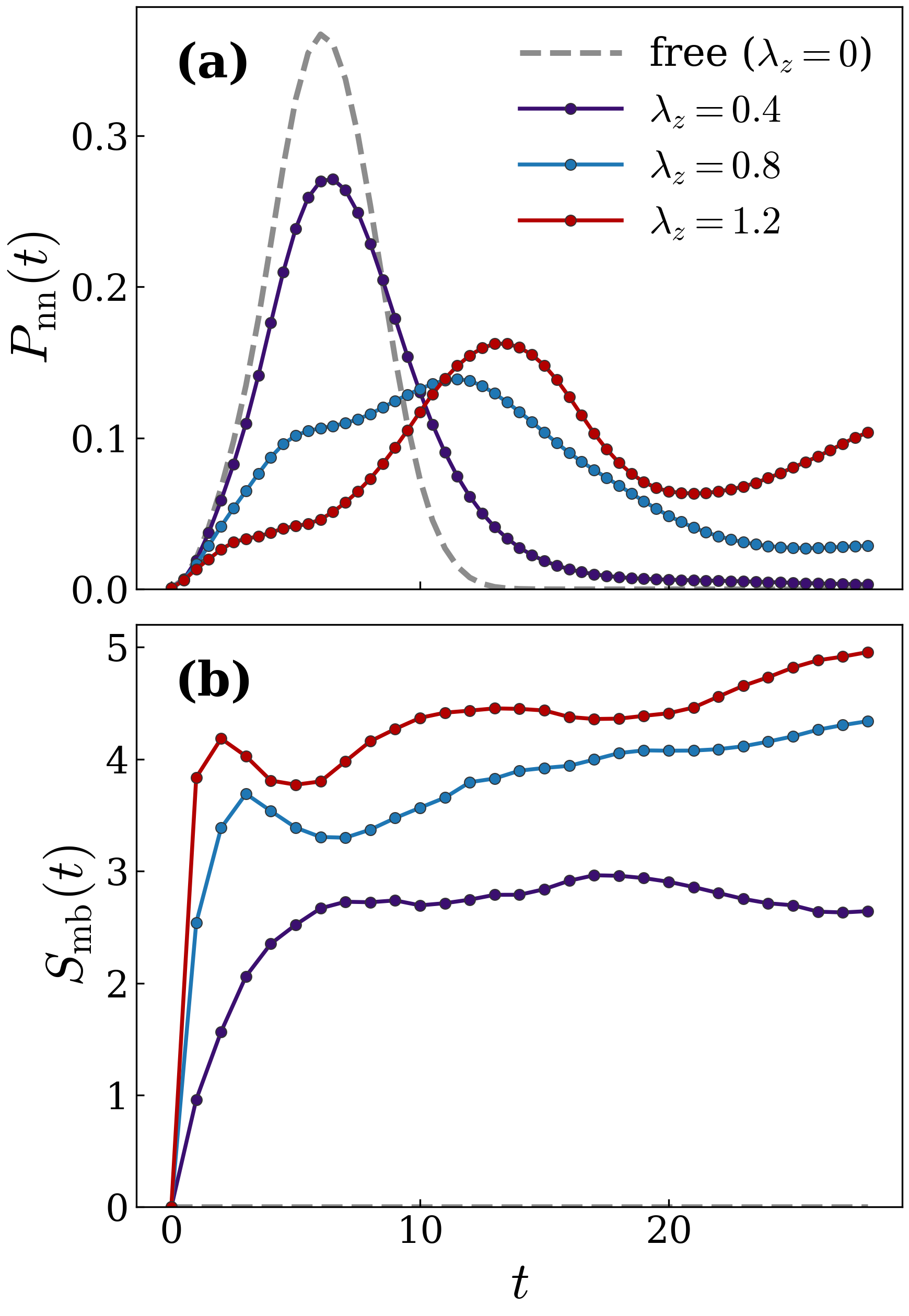}
  \caption{
  Real-time collision observables at $K=0$, with $t_{\mathrm{B}}=0.8$, $U=0$,
  $\omega_0=2$, $k_0=\pi/2$, and $L=32$, for
  $\lambda_z=0.4, 0.8, 1.2$ (colors) and the uncoupled reference
  $\lambda_z=0$ (dashed).
  (a) Adjacent-pair weight $P_{\mathrm{nn}}(t)=g(1,t)$. The transient maximum
  includes a kinematic contribution, whereas the post-collision value
  characterizes the retained compact component.
  (b) Magnon-boson entanglement entropy $S_{\mathrm{mb}}(t)$ across the
  magnon-boson partition.
  }
  \label{fig:pnnsmm}
\end{figure}

The time dependence of the local pairing weight and the magnon-boson entanglement is shown in Fig.~\ref{fig:pnnsmm}. The adjacent-pair weight $P_{\mathrm{nn}}(t)=g(1,t)$ increases as the wavepackets overlap. Its transient maximum is not, by itself, evidence of binding: for $k_0=\pi/2$, the free relative wavefunction $\sin(\pi r/2)$ already concentrates probability on odd separations. The value retained after the collision is therefore the more informative indicator of a compact component. The magnon-boson entropy $S_{\mathrm{mb}}(t)$ also increases as the initially bare magnons dress the bath during their approach and collision. As in the single-magnon problem of Sec.~\ref{sec:single}, this entropy quantifies magnon-boson correlations rather than pair binding alone.

\subsection{Spatial spin entanglement}

\begin{figure}[t]
  \centering
  \includegraphics[
    width=\columnwidth
  ]{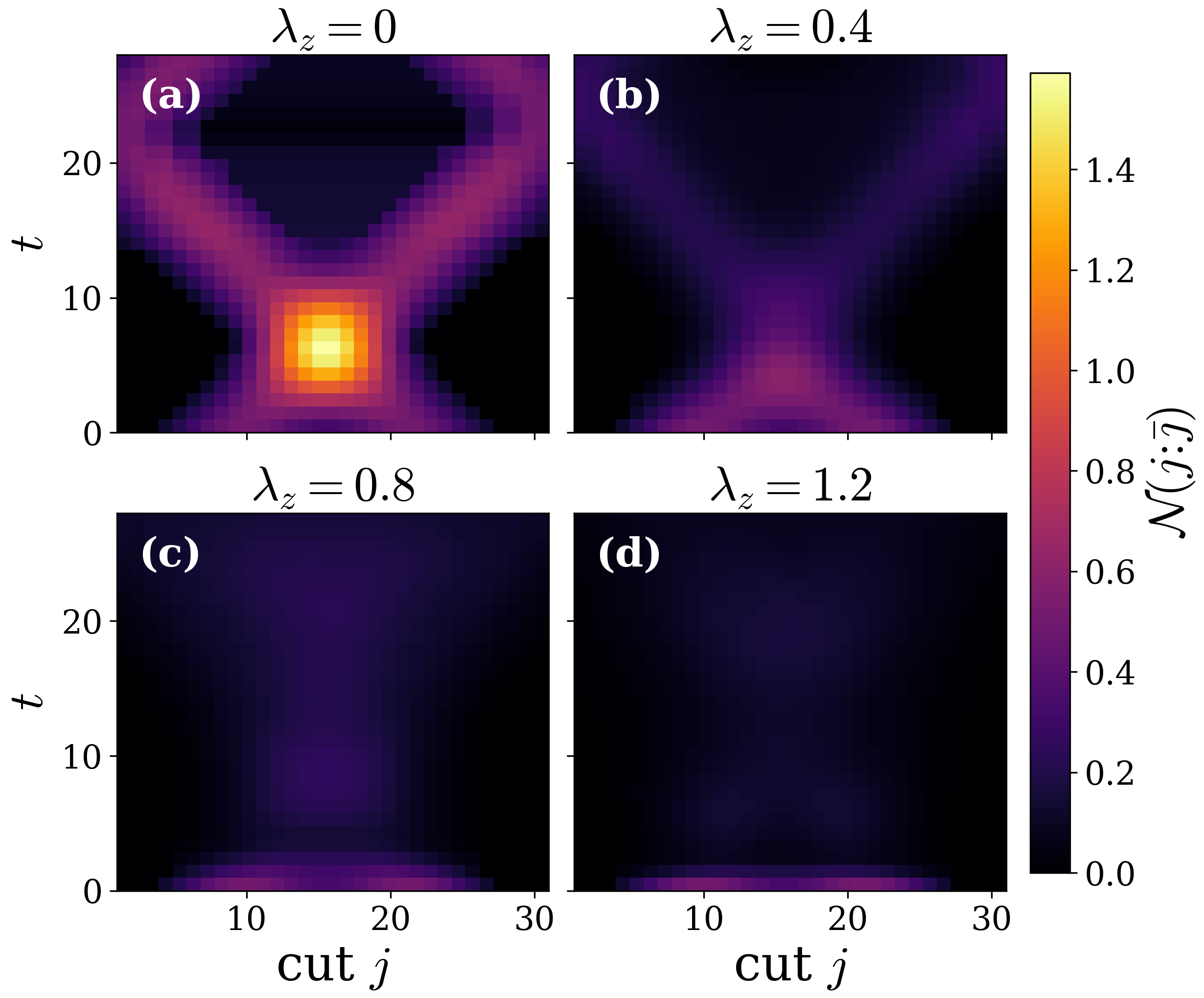}
  \caption{
  Spatiotemporal negativity $\mathcal{N}(j:\bar\jmath)$ of the reduced spin
  state across each cut $j$ during the collision. Panel (a) shows uncoupled
  magnons ($\lambda_z=0$), while panels (b-d) show increasing bath coupling.
  The extended negativity generated by the uncoupled collision is progressively
  suppressed with increasing $\lambda_z$. Parameters are $\omega_0=2$,
  $t_{\mathrm{B}}=0.8$, $U=0$, $L=32$, and initial wavepacket momenta $\pm k_0$,
  with $k_0=\pi/2$.
  }
  \label{fig:negprof}
\end{figure}

The entropy considered above measures entanglement across the magnon-boson bipartition. We now study a different correlation: the spatial entanglement that remains within the spin chain after the bath has been traced out. For the reduced spin state $\rho_{\mathrm{s}}$, we compute the negativity across each spatial cut $j$,
\begin{equation}
  \mathcal{N}(j:\bar\jmath)
  =
  \frac{
  \left\lVert\rho_{\mathrm{s}}^{T_{\mathcal L}}\right\rVert_1-1
  }{2},
\end{equation}
where $\rho_{\mathrm{s}}^{T_{\mathcal L}}$ denotes the partial transpose with respect to the left subchain $\mathcal L$ and $\lVert\cdot\rVert_1$ is the trace norm~\cite{VidalWerner2002}. Figure~\ref{fig:negprof} shows that the negativity is strongly suppressed as $\lambda_z$ increases.
In contrast, the uncoupled collision produces an extended pattern of spatial spin entanglement.

Two mechanisms can contribute to this suppression, and the present data show only their combined
effect. First, the formation of a compact and increasingly heavy pair restricts the relative
separation and spatial spreading of the magnons. Compactness alone does not necessarily eliminate
spatial entanglement, because a coherently delocalized composite object can still be entangled
across a cut. Second, different spin configurations become correlated with distinguishable
bosonic deformations. Their overlap is reduced on the local dressing scale, $\langle\alpha|\alpha'\rangle\sim e^{-\kappa^2}$, which suppresses off-diagonal coherence in $\rho_{\mathrm{s}}$ after the bath is traced out. The observed reduction of the negativity is therefore consistent with the combined influence of restricted pair motion and bath-induced loss of spin
coherence.

\section{Discussion and conclusions}
\label{sec:disc}

In this work, we have studied the real-time dynamics of one and two magnons coupled to a Bose-Hubbard bath using a semiclassical variational treatment and MPS-based simulations. By comparing these complementary approaches, we assess the role of quantum correlations beyond the semiclassical description.

For a single magnon, we have shown that the residual slowdown caused by retardation and the entanglement with the bath are governed by the same dressing weight once the static contribution is accounted for. The delayed bath deformation both slows the polaron and generates correlations beyond the position-dependent static dressing, causing the two effects to grow together. Because the magnon velocity is comparatively accessible, this relation enables transport to serve as a calibrated proxy for entanglement that would otherwise require knowledge of the full joint state. This correspondence is quantitatively controlled in the weak-dressing limit and remains approximately valid, up to a slowly varying logarithmic correction, within the off-resonant mobile regime explored here.

For two magnons, the bath mediates a finite-range attraction whose nearest-neighbor, antiadiabatic limit can be expressed as an effective XXZ anisotropy. 
Our finite-size ground-state calculations provide evidence that this interaction can bind the magnons. Across the parameter range explored, the binding energy and magnon-boson entanglement show parallel trends.
In real-time scattering, the interaction produces a compact component that persists after the collision. This retained component provides a more informative signature of bath-mediated pairing than the transient enhancement of the adjacent-pair weight during the collision, which also contains a purely kinematic contribution.

The on-site bath interaction controls both binding and magnon-boson entanglement. Increasing $U$ squeezes the bath's number fluctuations and reduces its response, causing both the binding energy and the magnon–boson entanglement to decrease. The induced interaction is not completely quenched in the hard-core limit but instead approaches a finite fraction of its harmonic-bath value. Moreover, strong dressing can produce a compact yet increasingly heavy pair. A reduction in the binding energy must therefore be distinguished from both the complete disappearance of the induced attraction and the loss of pair mobility. Together, these results identify bath deformability as a common control parameter for magnon transport, entanglement, and pairing.

Several ingredients of the Hamiltonian in Eq.~\eqref{eq:H} have been individually demonstrated or proposed across a range of quantum-simulation platforms. Superconducting circuit-QED arrays can realize interacting bosonic modes through microwave resonators with transmon-induced Kerr nonlinearities~\cite{SchmidtKoch2013,Hartmann2016}. These modes could potentially be coupled longitudinally to an adjacent qubit chain. Alternatively, Rydberg-atom arrays~\cite{Plodzien2018,Kosior2023,Mallick2025} and trapped ions~\cite{Solano2012,Cirac2012,Gorman2018,Sun2025} provide theoretically proposed or experimentally demonstrated routes to controllable couplings between internal spin states and local vibrational modes. Although these platforms provide several of the required spin-boson couplings and bath controls, integrating these elements into a complete realization of the interacting spin-boson dynamics considered here remains an open experimental challenge.
\begin{acknowledgments}
M.P., A.N.C., and A.R.\ acknowledge RES resources provided by the Barcelona Supercomputing Center on MareNostrum 5 under allocation NNO-2025-3-0004. This work has been partially funded by the Eric \& Wendy Schmidt Fund for Strategic Innovation through the CERN Next Generation Triggers project under grant agreement number SIF-2023-004.
\end{acknowledgments}

\textbf{Data Availability:} Data are available from the authors upon reasonable request.

\bibliographystyle{apsrev4-2}
\bibliography{refs_v2}

\appendix
\onecolumngrid

\pagebreak
\section{Lang-Firsov transformation}
\label{app:LF}

We transform with $\hat U=e^{-\hat S}$,
$\hat S=\kappa\sum_j(\ad_j-\hat a_j)\nm_j$, via the
Hadamard identity $\hat U\hat O\hat U^\dagger
=\hat O-[\hat S,\hat O]
+\tfrac1{2!}[\hat S,[\hat S,\hat O]]-\cdots$.
From $[\hat a_j,\ad_k]=\delta_{jk}$ one has
$[\ad_j,\hat n_j]=-\ad_j$ and
$[\hat a_j,\hat n_j]=\hat a_j$. Hence, the relevant building blocks are
\begin{equation}
\label{eq:bos}
  [\ad_j-\hat a_j,\hat n_j]
  =-(\ad_j+\hat a_j),
  \quad
  [\ad_j-\hat a_j,\ad_j+\hat a_j]
  =-2,
  \quad
  [\ad_j-\hat a_j,\hat a_j]
  =
  [\ad_j-\hat a_j,\ad_j]
  =-1.
\end{equation}

For the bosonic annihilation operator, the relevant commutator is
\[
[\hat S,\hat a_j]
=
\kappa\nm_j[\ad_j-\hat a_j,\hat a_j]
=
-\kappa\nm_j,
\]
and, since $[\hat S,\nm_j]=0$, the series truncates after the
first commutator,
\begin{equation}
\label{eq:bshift}
  \hat U\hat a_j\hat U^\dagger
  =
  \hat a_j+\kappa\nm_j
  \equiv\hat b_j,
  \qquad
  \hat U\ad_j\hat U^\dagger
  =
  \ad_j+\kappa\nm_j
  \equiv\hat b_j^\dagger .
\end{equation}

For the exchange, using
$[\nm_i,\Sp_j]=-\delta_{ij}\Sp_j$
($\Sp$ removes a magnon), we obtain
\[
[\hat S,\Sp_j]
=
-\kappa(\ad_j-\hat a_j)\Sp_j.
\]
The bosonic prefactor is invariant because
\[
[\hat S,\ad_j-\hat a_j]
=
\kappa\nm_j
[\ad_j-\hat a_j,\ad_j-\hat a_j]
=
0.
\]
The $n$-fold nested commutator is therefore
$(-\kappa)^n(\ad_j-\hat a_j)^n\Sp_j$, and the series resums to
\begin{equation}
\label{eq:Xdress}
  \hat U\Sp_j\hat U^\dagger
  =
  \sum_{n\geq0}
  \frac{\kappa^n}{n!}
  (\ad_j-\hat a_j)^n\Sp_j
  =
  \X_j\Sp_j,
  \qquad
  \X_j=e^{\kappa(\ad_j-\hat a_j)}.
\end{equation}
Likewise,
$\hat U\Sm_j\hat U^\dagger=\X_j^\dagger\Sm_j$,
so the exchange operator on a bond becomes
\[
\Sp_j\Sm_{j+1}\X_j\X_{j+1}^\dagger.
\]

For the on-site bath energy and the spin-bath coupling, using
$[\ad_j-\hat a_j,\hat n_j]=-(\ad_j+\hat a_j)$, we obtain
\begin{equation}
  \left[
    \hat S,
    \omega_0\sum_j\hat n_j
  \right]
  =
  -\omega_0\kappa
  \sum_j(\ad_j+\hat a_j)\nm_j
  =
  -\frac{\omega_0\kappa}{\lambda_z}
  \hat H_{\mathrm{int}},
\end{equation}
whereas
\begin{equation}
  [\hat S,\hat H_{\mathrm{int}}]
  =
  \kappa\lambda_z
  \sum_j(\nm_j)^2
  [\ad_j-\hat a_j,\ad_j+\hat a_j]
  =
  -2\kappa\lambda_z\sum_j\nm_j.
\end{equation}
Here, we used the projector property $(\nm_j)^2=\nm_j$. The series for the on-site bath energy terminates at second order, while that for the coupling terminates after the first commutator:
\begin{align}
  \hat U
  \left(
    \omega_0\sum_j\hat n_j
  \right)
  \hat U^\dagger
  &=
  \omega_0\sum_j\hat n_j
  +
  \frac{\omega_0\kappa}{\lambda_z}
  \hat H_{\mathrm{int}}
  +
  \kappa^2\omega_0\sum_j\nm_j,
  \\
  \hat U\hat H_{\mathrm{int}}\hat U^\dagger
  &=
  \hat H_{\mathrm{int}}
  +
  2\kappa\lambda_z\sum_j\nm_j.
\end{align}
The two terms linear in the bosonic operators combine into
\[
\left(
  1+\frac{\omega_0\kappa}{\lambda_z}
\right)
\hat H_{\mathrm{int}}.
\]
\newpage
They therefore vanish for

\begin{equation}
  \kappa=-\frac{\lambda_z}{\omega_0}.
\end{equation}
The remaining density term is
\begin{equation}
  \left(
    \kappa^2\omega_0+2\kappa\lambda_z
  \right)
  \sum_j\nm_j
  =
  -\frac{\lambda_z^2}{\omega_0}
  \sum_j\nm_j.
\end{equation}

This is the polaron self-energy appearing in Eq.~\eqref{eq:HLF}. It is constant within each sector of fixed
magnon number. For the bath hopping and the Bose-Hubbard interaction, using $\hat U\hat a_j\hat U^\dagger=\hat b_j$, one obtains
\begin{equation}
\label{eq:tBexp}
\begin{split}
  \hat U
  \left(
    -t_{\mathrm{B}}\ad_j\hat a_{j+1}
  \right)
  \hat U^\dagger
  =
  -t_{\mathrm{B}}\Big[
    &\ad_j\hat a_{j+1}
    +
    \kappa
    \left(
      \ad_j\nm_{j+1}
      +
      \nm_j\hat a_{j+1}
    \right) +
    \kappa^2\nm_j\nm_{j+1}
  \Big].
\end{split}
\end{equation}
Adding the Hermitian conjugate produces the density-density term twice on every bond:
\begin{equation}
\label{eq:VnnLF}
  -2\kappa^2t_{\mathrm{B}}
  \sum_j\nm_j\nm_{j+1}.
\end{equation}
This is the explicit nearest-neighbor contribution in
Eq.~\eqref{eq:Vbind} to the bath-mediated interaction in the
transformed Hamiltonian. It is not, by itself, the complete
induced interaction at finite $t_{\mathrm{B}}/\omega_0$. The transformation also produces the residual linear term
\begin{equation}
\label{eq:HresLF}
  -t_{\mathrm{B}}\kappa
  \sum_j
  \left(
    \ad_j\nm_{j+1}
    +
    \nm_j\hat a_{j+1}
    +
    \mathrm{H.c.}
  \right),
\end{equation}
which cannot be removed by the local generator. When the
remaining bath fluctuations are eliminated, this term produces
additional retarded and finite-range contributions. It underlies the retardation physics of Sec.~\ref{sec:retard}, contributes to the nonlinear potential derived in Appendix~\ref{app:nls}, and, together with Eq.~\eqref{eq:VnnLF}, recovers the complete static
potential derived in Appendix~\ref{app:BO}. Equation~\eqref{eq:tBexp} is exact. Since the displacement
in Eq.~\eqref{eq:bshift} is linear in $\kappa$, the expansion of each
quadratic bath-hopping operator terminates exactly at order
$\kappa^2$; no higher-order terms have been neglected. Finally, unitary conjugation preserves operator products, so
\begin{equation}
  \hat U\hat n_j\hat U^\dagger
  =
  \hat b_j^\dagger\hat b_j,
  \qquad
  \hat U
  \left[
    \frac{U}{2}\hat n_j(\hat n_j-1)
  \right]
  \hat U^\dagger
  =
  \frac{U}{2}
  \hat b_j^\dagger\hat b_j
  \left(
    \hat b_j^\dagger\hat b_j-1
  \right).
\end{equation}
Collecting all contributions yields
Eqs.~\eqref{eq:HLF}-\eqref{eq:HLFterms}.
\section{Davydov averages and equations of motion}
\label{app:dav}

Using the normal-ordering identity
\[
e^{\eta(\ad-\hat a)}
=
e^{\eta\ad}e^{-\eta\hat a}e^{-\eta^2/2},
\]
which is valid because $[\ad,\hat a]$ is a $c$-number, together with $\hat a\ket{\beta}=\beta\ket{\beta}$, we obtain
\begin{equation}
\label{eq:cohavg}
  \langle\beta|
  e^{\eta(\ad-\hat a)}
  |\beta\rangle
  =
  e^{-\eta^2/2}e^{\eta(\beta^*-\beta)}
  =
  e^{-\eta^2/2}e^{-2i\eta\Imm\beta}.
\end{equation}
With $\eta=+\kappa$ for $\X_j$ and $\eta=-\kappa$ for
$\X^\dagger_{j+1}$,
\begin{equation}
\label{eq:bondavg}
  \langle\X_j\rangle
  \langle\X^\dagger_{j+1}\rangle
  =
  e^{-\kappa^2}
  e^{-i(\theta_j-\theta_{j+1})},
  \qquad
  \theta_j=2\kappa\Imm\beta_j.
\end{equation}
The modulus gives the Lang-Firsov band narrowing, while the relative phase acts as a dynamical Peierls phase. For the shifted operator
\[
\hat b_j=\hat a_j+\kappa\nm_j,
\]
the same field $\{\beta_j\}$ accompanies every spin configuration in the single-field Davydov ansatz. Using the projector identity $(\nm_j)^2=\nm_j$, we obtain the exact variational averages
\begin{equation}
\label{eq:bexact}
\begin{split}
  \langle\hat b^\dagger_j\hat b_{j+1}\rangle
  &=
  \gamma_j^*\gamma_{j+1}
  +
  \kappa^2
  \langle\nm_j\nm_{j+1}\rangle_c,
  \\
  \left\langle
    \hat b^\dagger_j\hat b_j
    \left(
      \hat b^\dagger_j\hat b_j-1
    \right)
  \right\rangle
  &=
  \rho_j|\beta_j+\kappa|^4
  +
  (1-\rho_j)|\beta_j|^4,
\end{split}
\end{equation}
where
\begin{equation}
  \gamma_j=\beta_j+\kappa\rho_j,
  \qquad
  \rho_j=\langle\nm_j\rangle,
\end{equation}
and
\begin{equation}
  \langle\nm_j\nm_{j+1}\rangle_c
  =
  \langle\nm_j\nm_{j+1}\rangle
  -
  \rho_j\rho_{j+1}.
\end{equation}
Here, $\beta_j+\kappa$ and $\beta_j$ are the laboratory-frame fields conditioned on an occupied and an empty site, respectively. In the one-magnon sector,
\[
\langle\nm_j\nm_{j+1}\rangle=0
\]
because two different sites cannot be occupied simultaneously. Therefore,
\begin{equation}
  \langle\hat b^\dagger_j\hat b_{j+1}\rangle
  =
  \gamma_j^*\gamma_{j+1}
  -
  \kappa^2\rho_j\rho_{j+1}.
\end{equation}
The quartic average reduces to the Gross-Pitaevskii form $|\gamma_j|^4$ only when
\begin{equation}
  \kappa^2\rho_j(1-\rho_j)
\end{equation}
is negligible relative to $|\gamma_j|^2$. Replacing the conditioned quartic average by $|\gamma_j|^4$ constitutes an additional mean-field approximation, so we retain the conditioned expression in Eq.~\eqref{eq:bexact}. For a single magnon,
\begin{equation}
  \rho_j=|\psi_j|^2,
  \qquad
  \langle\nm_j\rangle=\rho_j,
  \qquad
  \langle\Sm_j\Sp_{j+1}\rangle
  =
  \psi_j^*\psi_{j+1}.
\end{equation}
Using Eq.~\eqref{eq:bexact}, the variational energy
$\mathcal E=\langle\Psi|\hat H_{\mathrm{LF}}|\Psi\rangle$ becomes
\begin{equation}
\label{eq:Efunc}
\begin{split}
  \mathcal E={}&
  -Je^{-\kappa^2}
  \sum_j
  \Ree\!\left[
    e^{i(\theta_j-\theta_{j+1})}
    \psi_j^*\psi_{j+1}
  \right]
  +
  \omega_0\sum_j|\beta_j|^2
  \\
  &-
  t_{\mathrm{B}}\sum_j
  \left(
    \gamma_j^*\gamma_{j+1}
    +
    \mathrm{c.c.}
    -
    2\kappa^2\rho_j\rho_{j+1}
  \right)
  \\
  &+
  \frac U2
  \sum_j
  \left[
    \rho_j|\beta_j+\kappa|^4
    +
    (1-\rho_j)|\beta_j|^4
  \right].
\end{split}
\end{equation}
The constant polaron self-energy
\[
-\frac{\lambda_z^2}{\omega_0}\sum_j\nm_j
\]
is omitted from Eq.~\eqref{eq:Efunc}, since it contributes only a global phase in the fixed one-magnon sector. The Davydov Lagrangian is
\begin{equation}
  \mathcal L
  =
  \langle\psi|i\partial_t|\psi\rangle
  +
  \sum_j
  \frac i2
  \left(
    \beta_j^*\dot\beta_j
    -
    \dot\beta_j^*\beta_j
  \right)
  -
  \mathcal E.
\end{equation}
Stationarity with respect to the independent complex variables gives
\begin{equation}
  i\dot\psi_j
  =
  \frac{\partial\mathcal E}{\partial\psi_j^*},
  \qquad
  i\dot\beta_j
  =
  \frac{\partial\mathcal E}{\partial\beta_j^*}.
\end{equation}
The required Wirtinger derivatives include
\begin{equation}
  \frac{\partial\theta_j}{\partial\beta_j^*}
  =
  i\kappa,
  \qquad
  \frac{\partial\gamma_k^*}{\partial\beta_j^*}
  =
  \delta_{jk},
  \qquad
  \frac{\partial\gamma_k}{\partial\psi_j^*}
  =
  \frac{\partial\gamma_k^*}{\partial\psi_j^*}
  =
  \kappa\psi_j\delta_{jk},
\end{equation}
together with their complex conjugates. These derivatives reproduce Eqs.~\eqref{eq:eom1}. We define
\[
\Phi_j=\frac{\partial\mathcal E}{\partial\rho_j},
\]
and find that the derivatives of the $\gamma$-dependent hopping combine with the counterterm in Eq.~\eqref{eq:Efunc}. Their neighboring-density contributions cancel, leaving
\begin{equation}
  -2t_{\mathrm{B}}\kappa
  \Ree\!\left(
    \beta_{j-1}+\beta_{j+1}
  \right),
\end{equation}
as in Eq.~\eqref{eq:Vj}; no constant force remains on the magnon-free background. The Bose-Hubbard term is linear in $\rho_j$ at fixed $\beta_j$ and gives
\begin{equation}
  \frac U2
  \left(
    |\beta_j+\kappa|^4
    -
    |\beta_j|^4
  \right).
\end{equation}
The final contribution is the current-divergence source
\begin{equation}
  i\kappa
  \left(
    \mathcal J_{j,j+1}
    -
    \mathcal J_{j-1,j}
  \right)
\end{equation}
in Eq.~\eqref{eq:eom1b}, which follows entirely from the dependence of the Peierls phases $\theta_j-\theta_{j+1}$ on $\beta_j^*$. The continuous variational flow conserves
\begin{equation}
  \sum_j|\psi_j|^2=1,
\end{equation}
because $\Phi_j$ is real and the hopping is Hermitian. Time-translation invariance also conserves $\mathcal E$; numerical preservation of both quantities depends on the integrator. Laboratory-frame observables are obtained from
\begin{equation}
  \langle\hat O\rangle_{\mathrm{lab}}
  =
  \langle\Psi|
  \hat U\hat O\hat U^\dagger
  |\Psi\rangle.
\end{equation}
The laboratory-frame bath amplitude is
\begin{equation}
  \langle\hat a_j\rangle_{\mathrm{lab}}
  =
  \beta_j+\kappa\rho_j
  \equiv\gamma_j,
\end{equation}
while the magnon density is frame independent,
\begin{equation}
  \langle\nm_j\rangle_{\mathrm{lab}}=\rho_j.
\end{equation}
The complete laboratory-frame boson occupation is
\begin{equation}
\label{eq:nlabdav}
  \langle\hat n_j\rangle_{\mathrm{lab}}
  =
  |\gamma_j|^2
  +
  \kappa^2\rho_j(1-\rho_j).
\end{equation}
Thus, $|\gamma_j|^2$ is only the coherent part of the cloud and equals the full occupation only when the fluctuation term is negligible. Since a magnon-free chain exerts no force, $\beta_j=0$ is its exact polaron-frame background and requires no density subtraction. However, with a magnon present, setting $\beta_j=0$ gives
\begin{equation}
  \gamma_j=\kappa\rho_j,
  \qquad
  \langle\hat n_j\rangle_{\mathrm{lab}}
  =
  \kappa^2\rho_j.
\end{equation}
This does not correspond to the bare laboratory-frame vacuum: for a delocalized magnon, the transformed vacuum is spin-conditioned and generally lies outside the single-field Davydov manifold.
\section{Nonlinear Schr\"odinger limit for a single magnon}
\label{app:nls}

The single-magnon equations \eqref{eq:eom1} reduce to a nonlinear Schr\"odinger equation when the bath is fast. We first take $U=0$ and assume $t_{\mathrm{B}}>0$, $\Omega_{\mathrm{B}}\equiv\omega_0-2t_{\mathrm{B}}>0$, and that the system is in the antiadiabatic regime
\begin{equation}
  \Omega_{\mathrm{mag}}\ll\Omega_{\mathrm{B}},
\end{equation}
where $\Omega_{\mathrm{mag}}$ is the characteristic frequency of the magnon packet. The bath then adjusts quickly enough to follow the magnon, so the time derivative of its field is negligible,
$|\dot\beta_j|\ll\Omega_{\mathrm{B}}|\beta_j|$. Dropping $\dot\beta_j$ in
\eqref{eq:eom1b} leaves an algebraic relation for the bath field,
\begin{equation}
\label{eq:nls_beta}
  (\omega_0-t_{\mathrm{B}}\hat\Delta)\beta_j
  =
  t_{\mathrm{B}}\kappa(\rho_{j-1}+\rho_{j+1})
  -i\kappa(\mathcal J_{j,j+1}-\mathcal J_{j-1,j}),
\end{equation}
with $\hat\Delta f_j\equiv f_{j-1}+f_{j+1}$. More precisely,
\[
  \omega_0-t_{\mathrm{B}}\hat\Delta
  =
  \Omega_{\mathrm{B}}+t_{\mathrm{B}}(2-\hat\Delta),
\]
where $2-\hat\Delta$ is the positive discrete lattice Laplacian. The full response operator is diagonal in Fourier space with eigenvalue $\omega(q)=\omega_0-2t_{\mathrm{B}}\cos q$. Its inverse therefore acts by convolution with the lattice Green's function $G_r$ of Eq.~\eqref{eq:Vr}, $(\omega_0-t_{\mathrm{B}}\hat\Delta)^{-1}f_j=\sum_kG_{j-k}f_k$. Applying this inverse to the
density-driven part of \eqref{eq:nls_beta}, neglecting the subleading current divergence for a broad, slowly moving packet, and restoring $\gamma_j=\beta_j+\kappa\rho_j$, we obtain
\begin{equation}
\label{eq:nls_defm}
  \gamma_j
  =
  \kappa\omega_0\sum_kG_{j-k}\rho_k
  +O(\Omega_{\mathrm{mag}}/\Omega_{\mathrm{B}}),
\end{equation}
where the background vanishes because of the density coupling. Thus, the magnon digs a cloud proportional to its own density. Substituting into the on-site potential \eqref{eq:Vj} gives a
self-interaction that is a convolution with the density. Its kernel is not the two-magnon potential $V(r)$. In the polaron frame the on-site dressing is already absorbed, and for a single hard-core magnon the Ising term of Eq.~\eqref{eq:Vbind} is inert ($\langle\nm_j\nm_{j+1}\rangle=0$). What remains is the second-order response to the residual bath-magnon source generated by the hopping, $-t_{\mathrm{B}}\kappa(\ad_j+\hat a_j)(\nm_{j-1}+\nm_{j+1})$ [Appendix~\ref{app:LF}], whose vertex carries the bond factor $2t_{\mathrm{B}}\kappa\cos q$,
\begin{equation}
\label{eq:nls_conv}
  \Phi_j
  =
  \sum_kV_{\mathrm{eff}}(j-k)\rho_k+{\mathrm{const}},
  \qquad
  \widehat V_{\mathrm{eff}}(q)
  =
  -\frac{8\kappa^2t_{\mathrm{B}}^2\cos^2q}
  {\omega_0-2t_{\mathrm{B}}\cos q},
\end{equation}
so that each magnon feels the potential generated by its own cloud. For a magnon with width $\ell$ much larger than both the lattice spacing and the response range, $\ell\gg a$ and $\ell\gg r_0a$, the convolution is local, $\Phi_j\simeq-g\rho_j$, with
\begin{equation}
\label{eq:nls_g}
  g
  =
  -\widehat V_{\mathrm{eff}}(0)
  =
  \frac{8\kappa^2t_{\mathrm{B}}^2}{\omega_0-2t_{\mathrm{B}}}
  =
  \left(\frac{2t_{\mathrm{B}}}{\omega_0}\right)^{\!2}
  \frac{2\lambda_z^2}{\omega_0-2t_{\mathrm{B}}},
\end{equation}
the $q=0$ weight of the effective potential. This expression grows toward the soft-bath edge $2t_{\mathrm{B}}\to\omega_0$, but it cannot be extrapolated to that limit: the response time $\Omega_{\mathrm{B}}^{-1}$ and the range $r_0$ diverge there,
invalidating both the adiabatic and the local approximations. The Bose-Hubbard nonlinearity weakens the self-focusing. Keeping the exact $U$ term of \eqref{eq:eom1b} in the slaved equation adds the conditioned-field source $-U\kappa^3\rho_j$ to \eqref{eq:nls_beta} and the linear response $2U\kappa^3\Ree\beta_j$ to the potential \eqref{eq:Vj};
both carry one factor of the cloud and one of $U$. To first order in $U\kappa^2/t_{\mathrm{B}}\ll1$, they rescale the cubic coupling,
\begin{equation}
\label{eq:nls_gU}
  g(U)
  =
  g\left[
    1-\frac{U\kappa^2}{t_{\mathrm{B}}}
    +O\!\left(\frac{U^2\kappa^4}{t_{\mathrm{B}}^2}\right)
  \right].
\end{equation}
The density-independent term gives only a uniform energy shift, while higher powers of $\rho_j$ generate quintic and higher nonlinearities, which are omitted in the cubic NLS limit. The interaction stiffens the bath against the magnon's imprint, thereby weakening the mean-field self-interaction. This weakening is the
single-magnon counterpart of the depth suppression $f(U)$ of the two-magnon potential [Eq.~\eqref{eq:VfU}] and has the same physical origin: $U$ caps the bath's response to the source. Because the correction enters at relative order $U\kappa^2/t_{\mathrm{B}}$, the single-magnon $U$ dependence
is parametrically weak at weak dressing. Equation~\eqref{eq:eom1a} is then a discrete focusing nonlinear
Schr\"odinger equation. For the continuum limit, we use
\[
  \psi_j=\sqrt a\,\psi(x),
  \qquad
  x=ja,
  \qquad
  \int dx\,|\psi(x)|^2=1.
\]
For an envelope centered near the bottom of the dressed band ($k_0\simeq0$), with the Peierls phase dropped and the uniform band-bottom energy removed, the dressed hopping expands to
\begin{equation}
\label{eq:nls_cont}
  i\partial_t\psi
  =
  -\frac{1}{2m^*}\partial_x^2\psi
  -g_{\mathrm{c}}|\psi|^2\psi,
  \qquad
  \frac{1}{2m^*}
  =
  \frac{\widetilde J a^2}{2},
  \qquad
  g_{\mathrm{c}}=ga,
  \qquad
  \widetilde J=Je^{-\kappa^2}.
\end{equation}
For a carrier momentum $k_0$, the quadratic envelope coefficient is instead
$1/[2m^*(k_0)]=\widetilde J a^2\cos(k_0)/2$. It vanishes at $k_0=\pi/2$, so the quadratic NLS and the soliton solution below do not apply to the retardation simulations performed at that momentum.
The normalized bright soliton is
\begin{equation}
  \psi(x,t)
  =
  \frac{1}{\sqrt{2\xi}}
  \operatorname{sech}\!\left(\frac{x-x_0}{\xi}\right)e^{-i\mu t},
  \qquad
  \xi=\frac{2}{m^*g_{\mathrm{c}}},
  \qquad
  \mu=-\frac{1}{2m^*\xi^2}.
\end{equation}
It describes a self-focused magnon only when $\xi\gg a$. When
$\xi$ approaches the lattice scale, the continuum and local approximations are no longer controlled. The self-trapping line $\max_j|\gamma_j|^2>1$ used in the main text is therefore an operational diagnostic of the full lattice mean-field dynamics, rather than a consequence of the continuum NLS. Restoring $\dot\beta_j$ recovers the full mean-field dynamics of \eqref{eq:eom1}, in which the classical cloud lags the magnon and slows it below $v_{\mathrm{LF}}$. That lag is classical. The Davydov ansatz represents the spin and bath as a product state in the polaron frame, with the bath described by a product of coherent states, so the polaron-frame entanglement $S_{\mathrm{mb}}^{\mathrm{LF}}=0$ identically, independently
of $\dot\beta_j$. In the laboratory frame the mean field still carries the deterministic adiabatic entanglement of Eq.~\eqref{eq:rho_adiabatic} (the $e^{-\kappa^2}$ suppression of position coherence), but, within the single-field manifold, the retardation lag
adds no further entanglement. The excess entanglement generated by the bath's finite quantum response and the part of the retardation carried by that response appear only when the bath is treated beyond a single coherent state, as in the MPS treatment of Sec.~\ref{sec:retard}.
\section{Weak-coupling relation between retardation and magnon-boson entanglement}
\label{app:collapse}

Consider a bare magnon wavepacket with $k_0=\pi/2$ released in the presence of an Einstein bath ($t_{\mathrm{B}}=0$)~\cite{KuTrugman2007}. For $\omega_0>J$, both the second-order velocity deficit and the leading entanglement are governed by the dressing weight $\bar n$, up to a slowly varying factor fixed by the momentum distribution and packet width. Eliminating $\bar n$ yields the relation in Fig.~\ref{fig:retard}. For driven and resonant polarons, see Refs.~\cite{Vidmar2011,Golez2012}. In the strict antiadiabatic limit, the coupling is diagonal in the magnon position and a pinned magnon realizes the independent-boson model~\cite{Mahan}. A magnon with position amplitudes $\phi_j$ then has the joint state
\begin{equation}
\label{eq:adiabatic_state}
  \ket{\Psi}
  =
  \sum_j\phi_j\,\ket{j}\otimes\hat D_j\ket{0},
\end{equation}
with
\[
  \hat D_j=e^{\kappa(\ad_j-\hat a_j)},
  \qquad
  \kappa=-\frac{\lambda_z}{\omega_0}.
\]
This is Eq.~\eqref{eq:ansatz1} in the laboratory frame at $\beta_k=0$; it is exact for a pinned magnon and asymptotically exact for an adiabatically dressed mobile magnon as $J/\omega_0\to0$. Since the Einstein modes are independent,
\[
  \bra{0}\hat D_j^\dagger\hat D_{j'}\ket{0}
  =
  e^{-\kappa^2},
  \qquad
  j\ne j',
\]
and tracing out the bath gives
\begin{equation}
\label{eq:rho_adiabatic}
  \rho_{jj'}
  =
  \phi_j\phi_{j'}^*
  e^{-\kappa^2(1-\delta_{jj'})}.
\end{equation}
The populations are unchanged, while intersite coherences are reduced by $e^{-\kappa^2}$. A localized magnon remains unentangled, whereas a spatial superposition correlates distinct positions with different bath states. The same overlap narrows the hopping to $\widetilde J=Je^{-\kappa^2}$. Equation~\eqref{eq:rho_adiabatic} is exact for the static state \eqref{eq:adiabatic_state}; the released bare packet requires a dynamical treatment. On a periodic chain, define
$\hat c_j^\dagger\equiv\Sm_j$ on the polarized background,
$\nm_j=\hat c_j^\dagger\hat c_j$, and
\[
  \hat c_k^\dagger
  =
  \frac{1}{\sqrt L}
  \sum_j e^{ikj}\hat c_j^\dagger,
\]
which requires no Jordan-Wigner string in the one-magnon sector. Fourier transformation gives
\begin{equation}
\label{eq:Hk}
  \hat H
  =
  \sum_k\varepsilon_k\hat c_k^\dagger\hat c_k
  +
  \omega_0\sum_q\ad_q\hat a_q
  +
  \frac{\lambda_z}{\sqrt L}
  \sum_{k,q}
  \left(
    \hat c_{k-q}^\dagger\hat c_k\ad_q
    +
    \mathrm{H.c.}
  \right),
  \qquad
  \varepsilon_k=-J\cos k.
\end{equation}
Locality makes the coupling matrix element momentum independent: a boson of momentum $q$ causes the magnon to recoil from $k$ to $k-q$ while conserving total momentum. At fixed $k_0$, the zero- and one-boson subspace is spanned by
\begin{equation}
\label{eq:basis}
  \ket{k_0;0}
  \quad\text{and}\quad
  \ket{k_0-q;1_q},
  \qquad
  q\in\mathrm{BZ}.
\end{equation}
The basis states describe the bare magnon and the recoiled magnon accompanied by one boson. Two-boson states enter the wavefunction at order $\lambda_z^2$ and the observables at order $\lambda_z^4$. Using interaction-picture amplitudes, we write
\begin{equation}
\label{eq:ansatzPT}
  \ket{\Psi(t)}
  =
  c_0(t)e^{-i\varepsilon_{k_0}t}\ket{k_0;0}
  +
  \sum_q
  c_q(t)
  e^{-i(\varepsilon_{k_0-q}+\omega_0)t}
  \ket{k_0-q;1_q}.
\end{equation}
The Schr\"odinger equation, with the initial conditions
$c_0(0)=1$ and $c_q(0)=0$, gives, to first order,
\[
  i\dot c_q
  =
  \frac{\lambda_z}{\sqrt L}e^{-i\Delta_qt},
\]
where the off-shell energy mismatch is
\begin{equation}
\label{eq:mismatch}
  \Delta_q
  =
  \varepsilon_{k_0}
  -
  \varepsilon_{k_0-q}
  -
  \omega_0.
\end{equation}
Therefore,
\begin{equation}
\label{eq:cq}
  c_q(t)
  =
  \frac{\lambda_z}{\sqrt L}
  \frac{e^{-i\Delta_qt}-1}{\Delta_q},
  \qquad
  |c_q(t)|^2
  =
  \frac{2\lambda_z^2}{L\Delta_q^2}
  \left(1-\cos\Delta_qt\right).
\end{equation}

For $k_0=\pi/2$ and $\omega_0>J$, the mismatch never vanishes:
\[
  \Delta_q\leq J-\omega_0<0.
\]
Thus, no boson is emitted on shell; the occupations remain bounded and oscillatory. Time averaging gives
\[
  \overline{|c_q|^2}
  =
  \frac{2\lambda_z^2}{L\Delta_q^2}.
\]
Their sum defines the time-averaged one-boson, or dressing, weight $\bar n$.

For $k_0=\pi/2$, one has
$\varepsilon_{k_0}=0$,
$\varepsilon_{k_0-q}=-J\sin q$, and hence
\[
  \Delta_q=J\sin q-\omega_0.
\]
The momentum sum becomes
\begin{equation}
\label{eq:nbar_int}
  \bar n
  =
  2\lambda_z^2
  \int_0^{2\pi}
  \frac{dq}{2\pi}
  \frac{1}{(\omega_0-J\sin q)^2}.
\end{equation}
Shifting $q\to q+\pi/2$ and using
\[
  \int_0^{2\pi}
  \frac{dq}{2\pi}
  \frac{1}{A-B\cos q}
  =
  \frac{1}{\sqrt{A^2-B^2}},
  \qquad
  A>|B|,
\]
and
\[
  \int_0^{2\pi}
  \frac{dq}{2\pi}
  \frac{1}{(A-B\cos q)^2}
  =
  \frac{A}{(A^2-B^2)^{3/2}},
\]
with $A=\omega_0$ and $B=J$, we obtain
\begin{equation}
\label{eq:nbar}
  \bar n
  \equiv
  \sum_q\overline{|c_q|^2}
  =
  \frac{2\kappa^2}
  {\left[1-(J/\omega_0)^2\right]^{3/2}},
  \qquad
  \kappa=-\frac{\lambda_z}{\omega_0}.
\end{equation}

The packet velocity is obtained from the current operator associated with the conserved magnon number,
\[
  \hat v
  =
  \sum_k
  \varepsilon'_k
  \hat c_k^\dagger\hat c_k,
\]
which is diagonal in the basis \eqref{eq:basis}; hence,
\begin{equation}
\label{eq:vexp}
  \langle\hat v\rangle
  =
  |c_0|^2\varepsilon'_{k_0}
  +
  \sum_q|c_q|^2\varepsilon'_{k_0-q}
  =
  \varepsilon'_{k_0}
  -
  \sum_q|c_q|^2
  \left(
    \varepsilon'_{k_0}
    -
    \varepsilon'_{k_0-q}
  \right).
\end{equation}
Using $|c_0|^2=1-\sum_q|c_q|^2$, time averaging at $k_0=\pi/2$ gives
\begin{equation}
\label{eq:vbar1}
  \bar v
  =
  J
  \left[
    1
    -
    2\lambda_z^2
    \int_0^{2\pi}
    \frac{dq}{2\pi}
    \frac{1-\cos q}
    {(\omega_0-J\sin q)^2}
  \right].
\end{equation}
The term proportional to $\cos q$ integrates to zero because
\[
  \frac{d}{dq}
  \left[
    \frac{1}{J(\omega_0-J\sin q)}
  \right]
  =
  \frac{\cos q}{(\omega_0-J\sin q)^2},
\]
and the bracket is periodic over the Brillouin zone. The remaining integral is precisely Eq.~\eqref{eq:nbar_int}, yielding the second-order velocity law
\begin{equation}
\label{eq:vlaw}
  \bar v
  =
  v_{\mathrm{bare}}(1-\bar n),
\end{equation}
where
$v_{\mathrm{bare}}=J\sin k_0=J$ at $k_0=\pi/2$.

The velocity deficit relative to the Lang-Firsov value follows by expanding
\[
  v_{\mathrm{LF}}
  =
  v_{\mathrm{bare}}e^{-\kappa^2}
  =
  v_{\mathrm{bare}}
  \left(
    1-\kappa^2
  \right)
  +
  O(\kappa^4).
\]
Consequently,
\[
  \frac{\bar v}{v_{\mathrm{LF}}}
  =
  \frac{1-\bar n}{1-\kappa^2}
  =
  1-(\bar n-\kappa^2)+O(\kappa^4),
\]
and
\begin{equation}
\label{eq:deltav}
  \delta v
  \equiv
  1-\frac{\bar v}{v_{\mathrm{LF}}}
  =
  \bar n-\kappa^2
  =
  \kappa^2
  \left[
    \frac{2}
    {\left[1-(J/\omega_0)^2\right]^{3/2}}
    -
    1
  \right].
\end{equation}
Equation~\eqref{eq:deltav} has no fitted parameter and agrees with the offset-corrected tensor-network deficits of Appendix~\ref{app:num} throughout the weak-dressing regime; corrections enter at relative order $\bar n$. For a plane wave, the bath states are orthogonal:
\[
  \langle0|1_q\rangle=0,
  \qquad
  \langle1_q|1_{q'}\rangle=\delta_{qq'}.
\]
Up to the negligible $q=0$ degeneracy at finite $L$, tracing out the bath therefore leaves a magnon density matrix diagonal in momentum,
\begin{equation}
\label{eq:rhom}
  \hat\rho_{\mathrm{m}}(t)
  =
  |c_0(t)|^2\ket{k_0}\!\bra{k_0}
  +
  \sum_q
  |c_q(t)|^2
  \ket{k_0-q}\!\bra{k_0-q},
\end{equation}
with instantaneous entropy
\begin{equation}
  S_{\mathrm{mb}}(t)
  =
  -|c_0(t)|^2\ln|c_0(t)|^2
  -
  \sum_q|c_q(t)|^2\ln|c_q(t)|^2.
\end{equation}

A plane wave produces a $\ln L$ contribution because the recoiled momenta are orthogonal. This expression does not apply directly to the finite packet used numerically: for width $\sigma$, recoils within the momentum resolution $1/\sigma$ overlap, replacing the system-size dependence by a packet-width dependence. Writing the plane-wave weights as
\[
  \overline{|c_q|^2}
  =
  \bar n\,p_q,
  \qquad
  \sum_q p_q=1,
\]
the entropy of the dephased reduced state constructed from these time-averaged weights is
\begin{equation}
\label{eq:Sexpand}
\begin{split}
  S_{\mathrm{mb}}^{\mathrm{deph}}
  &=
  -(1-\bar n)\ln(1-\bar n)
  -
  \sum_q
  \bar n p_q\ln(\bar n p_q)
  \\
  &=
  \bar n
  \left[
    (1+H_p)-\ln\bar n
  \right]
  +
  O(\bar n^2),
\end{split}
\end{equation}
where
\[
  H_p=-\sum_q p_q\ln p_q.
\]
Because the von Neumann entropy is nonlinear, the entropy of the time-averaged state in Eq.~\eqref{eq:Sexpand} is not identical to the time average of the instantaneous entropy. Their leading weak-coupling structure is nevertheless the same:
\begin{equation}
\label{eq:Sstruct}
  S_{\mathrm{mb}}^*
  =
  \bar n
  \left[
    \mathcal A
    -
    \mathcal B\ln\bar n
  \right]
  +
  O(\bar n^2).
\end{equation}
For a plane wave in the thermodynamic limit, $\mathcal B=1$, while $\mathcal A$ contains the normalized momentum entropy and a coupling-independent temporal factor associated with averaging the instantaneous entropy. Packet overlaps modify both coefficients. For the broad packet used here, these corrections are small and $\mathcal B\simeq1$, while $\mathcal A$ retains a weak dependence on the band, bath frequency, and resolution $1/\sigma$.

For the dashed curves in Fig.~\ref{fig:retard}, no assumption of an exactly diagonal packet density matrix is made. The amplitudes \eqref{eq:cq} are superposed over the simulated momentum distribution, the reduced density matrix is constructed and diagonalized, and its entropy is averaged over the numerical time window. No fit is used; agreement is good for $\omega_0\gtrsim4J$ and deteriorates near $\omega_0=J$.

After sudden release, the initially undressed magnon builds a cloud that oscillates about its displaced configuration, making the time-averaged laboratory-frame excitation weight twice the stationary value in the antiadiabatic limit. This does not imply an exact entropy ratio because of the logarithmic dependence; such a ratio is approached only at leading logarithmic order when the branch distributions are comparable. Equations~\eqref{eq:vlaw} and \eqref{eq:Sstruct} instead share the same dominant weight $\bar n$, while $\mathcal A$ and $\mathcal B$ retain the residual bath-frequency and packet-width dependence. Eliminating
\[
  \bar n=\delta v+\kappa^2
\]
gives
\begin{equation}
\label{eq:SofdV}
  S_{\mathrm{mb}}^*
  =
  \left(
    \delta v+\kappa^2
  \right)
  \left[
    \mathcal A
    -
    \mathcal B
    \ln\left(
      \delta v+\kappa^2
    \right)
  \right]
  +
  O\!\left[
    \left(
      \delta v+\kappa^2
    \right)^2
  \right].
\end{equation}
For a fixed packet, the velocity deficit is therefore a weak-coupling proxy for the entanglement, with the residual dependence encoded in $\mathcal A$ and $\mathcal B$. The derivation assumes $k_0=\pi/2$ and $\omega_0>J$. Below this threshold, some momenta satisfy $\Delta_q=0$, allowing real bath emission and invalidating Eq.~\eqref{eq:deltav}.
\section{Two-magnon sector and bound state}\label{app:two}

Unlike the single-magnon problem, whose antiadiabatic mean-field dynamics admits a focusing nonlinear Schr\"odinger equation with a bright-soliton solution (Appendix~\ref{app:nls}), the mutual binding of two hard-core magnons, which cannot occupy the same lattice site, is instead described by a linear two-body problem. For the full finite-range induced potential, the size of the resulting lattice molecule depends jointly on the range $r_0$, the potential depth, and the relative kinetic energy. In the
nearest-neighbor antiadiabatic reduction considered below, the decay length of the bound state is set by the ratio of $\Vb$ to the relative hopping, not by a continuum soliton width.

Within this nearest-neighbor effective model, the explicit pair interaction is kept at the operator level rather than factorized. The pair term $-\Vb\sum_j\nm_j\nm_{j+1}$ in Eq.~\eqref{eq:Vbind} acts on $\psi_{jk}$ with matrix element $-\Vb\,\delta_{|j-k|,1}$, while the pair correlation is
$\langle\nm_j\nm_{j+1}\rangle=|\psi_{j,j+1}|^2$. Consequently, the $t_{\mathrm{B}}$ part of the potential $\Phi_j$ must be written in terms of $\beta_j$, rather than solely in terms of $\gamma_j$ [Eq.~\eqref{eq:Vj}]: using the $\gamma$ form without the connected-correlation correction would add the mean-field density product $\kappa^2\rho_j\rho_{j+1}$ and double-count the explicit interaction.

For the nearest-neighbor antiadiabatic bound-state estimate, projecting the remaining bath fluctuations onto the polaron-frame vacuum ($\beta_j\equiv0$, $U=0$) leaves hard-core magnons
with hopping $\widetilde J=Je^{-\kappa^2}$ and attraction $\Vb$. Translational invariance allows the center-of-mass motion to be separated as
$\psi_{jk}=e^{iK(j+k)/2}\phi(r)$, with $r=k-j\ge1$. Taking
$K\in[-\pi,\pi]$, the relative amplitude obeys, for $r\ge2$,
\begin{equation}\label{eq:rel}
  E\,\phi(r)
  =
  -\tau\big[\phi(r+1)+\phi(r-1)\big],
  \qquad
  \tau=\widetilde J\cos(K/2)\ge0,
\end{equation}
with the hard-core boundary $\phi(0)=0$ and, at $r=1$,
\[
  E\phi(1)=-\tau\phi(2)-\Vb\phi(1).
\]
A bound state is an exponentially decaying solution
$\phi(r)=\mathcal C z^{r-1}$ with $|z|<1$. Substitution into the bulk equation gives $E=-\tau(z+z^{-1})$, whereas the boundary equation gives $E=-\tau z-\Vb$. Equating the two expressions,
\[
  -\tau(z+z^{-1})=-\tau z-\Vb,
\]
yields $\tau z^{-1}=\Vb$ and therefore
\begin{equation}\label{eq:zsol}
  z=\frac{\tau}{\Vb},
  \qquad
  |z|<1
  \ \Leftrightarrow\
  \Vb>\tau
  =
  Je^{-\kappa^2}\cos(K/2),
\end{equation}
which is the threshold $|\Deff|>\cos(K/2)$ of Eq.~\eqref{eq:thr}. For $0<\tau<\Vb$, the relative-wavefunction decay length, in lattice units, is
\begin{equation}
  \xi_{\mathrm{rel}}
  =
  \frac{1}{\ln(\Vb/\tau)}.
\end{equation}
At $\tau=0$, the pair is strictly confined to $r=1$. The bound-state energy within the reduced relative problem is
\[
  E_{\mathrm{pair}}(K)
  =
  -\tau(z+z^{-1})
  =
  -\left(\frac{\tau^2}{\Vb}+\Vb\right).
\]
The two-magnon continuum bottom at fixed total momentum $K$ is $-2\tau$. Hence, the positive binding energy, defined as the energy by which the bound state lies below the continuum, is
\[
  E_b
  =
  -2\tau-E_{\mathrm{pair}}(K)
  =
  -2\tau+\frac{\tau^2}{\Vb}+\Vb
  =
  \frac{(\Vb-\tau)^2}{\Vb}.
\]
When the full finite-range potential $V(r)$ of Eq.~\eqref{eq:Vr} is retained, the term $-\Vb\delta_{r,1}$ is replaced by $V(r)$, and the bound-state profile and threshold must in general be determined from the corresponding discrete Schr\"odinger equation.
\section{Born-Oppenheimer bath-mediated potential}\label{app:BO}

We derive Eq.~\eqref{eq:Vr} for noninteracting bosons ($U=0$). In the heavy-magnon limit
$J\to0$, the magnon numbers $\nm_j$ are conserved and the spins are static. For a fixed magnon
configuration, each operator $\nm_j$ can therefore be replaced by its eigenvalue $n_j^{\mathrm{m}}\in\{0,1\}$, so the coupling
$\hat H_{\mathrm{int}}=\lambda_z\sum_j(\ad_j+\hat a_j)\nm_j$ acts as a set of classical linear sources for the bath. The corresponding source-conditioned bath Hamiltonian contains quadratic and linear terms,
\begin{equation}\label{eq:HBsource}
  \hat H_{\mathrm{BO}}=\omega_0\sum_j\ad_j\hat a_j-t_{\mathrm{B}}\sum_j(\ad_j\hat a_{j+1}+\mathrm{H.c.})
   +\sum_j c_j(\hat a_j+\ad_j),\qquad c_j=\lambda_z n_j^{\mathrm{m}} .
\end{equation}

In Fourier space, with $\hat a_j=L^{-1/2}\sum_q e^{iqj}\hat a_q$ for a periodic bath of $L$ sites, the quadratic part is diagonal,
\begin{equation}\label{eq:HBfree}
  \omega_0\sum_j\ad_j\hat a_j-t_{\mathrm{B}}\sum_j(\ad_j\hat a_{j+1}+\mathrm{H.c.})
   =\sum_q\omega(q)\,\ad_q\hat a_q,\qquad \omega(q)=\omega_0-2t_{\mathrm{B}}\cos q.
\end{equation}
For the convention $t_{\mathrm{B}}\ge0$ used here, the bath is positive and gapped when $2t_{\mathrm{B}}<\omega_0$. The linear source becomes
\begin{equation}\label{eq:source}
  \sum_j c_j(\hat a_j+\ad_j)=\sum_q\big(C_q\,\ad_q+C_q^*\,\hat a_q\big),\qquad
  C_q=L^{-1/2}\sum_j c_j\,e^{-iqj}.
\end{equation}
The modes decouple, $\hat H_{\mathrm{BO}}=\sum_q\hat h_q$, with
$\hat h_q=\omega(q)\ad_q\hat a_q+C_q\ad_q+C_q^*\hat a_q$. Completing the square in each mode with the $c$-number shift
$\hat a_q=\tilde a_q-C_q/\omega(q)$ gives
\begin{equation}\label{eq:complete}
\begin{split}
  \hat h_q&=\omega(q)\Big(\ad_q+\tfrac{C_q^*}{\omega(q)}\Big)
  \Big(\hat a_q+\tfrac{C_q}{\omega(q)}\Big)
   -\frac{|C_q|^2}{\omega(q)}\\
   &=\omega(q)\,\tilde a_q^\dagger\tilde a_q-\frac{|C_q|^2}{\omega(q)} ,
\end{split}
\end{equation}
a harmonic oscillator displaced by $-C_q/\omega(q)$ with its energy lowered by
$|C_q|^2/\omega(q)$. The ground-state energy is the sum of these shifts,
\begin{equation}\label{eq:E0q}
  E=-\sum_q\frac{|C_q|^2}{\omega(q)} .
\end{equation}

Inserting $|C_q|^2=L^{-1}\sum_{j,k}c_jc_k\,e^{-iq(j-k)}$ for real $c_j$
into Eq.~\eqref{eq:E0q} and summing over $q$ first gives
\begin{equation}\label{eq:Ereal}
  E=-\sum_{j,k}c_jc_k\,G_{j-k},\qquad
  G_r\equiv\frac1L\sum_q\frac{e^{-iqr}}{\omega(q)}
  =\frac1L\sum_q\frac{\cos(qr)}{\omega(q)},
\end{equation}
with $G_r$ real and even since $\omega(-q)=\omega(q)$. The energy $E$ in Eq.~\eqref{eq:Ereal} is a quadratic form in the sources $\{c_j\}$, so the two-body interaction is isolated by a subtraction. Adding a magnon at an unoccupied site $m$ switches on the source there, changing $c_m$ by $\delta c_m=+\lambda_z$. For two hard-core magnons at sites $0$ and $r\ge1$, their interaction
is the mixed second difference of $E$, which retains only the cross term,
\begin{equation}\label{eq:Vsub}
  V(r)=E(\{0,r\})-E(\{0\})-E(\{r\})+E(\varnothing)
   =-2\,\delta c_0\,\delta c_r\,G_r=-2\lambda_z^2\,G_r ,
\end{equation}
with the background energy and the two individual self-energies cancelling. Thus, the interaction
is determined by the coupling squared and the static bath Green's function $G_r$ and is the lattice
analogue of the bus-mediated spin-spin interaction of Ref.~\cite{Cirac2004}. In the thermodynamic limit $L\to\infty$, the lattice Green's function is
\begin{equation}\label{eq:Gint}
  G_r=\frac1{2\pi}\int_0^{2\pi}\!\frac{\cos(rq)}{\omega_0-2t_{\mathrm{B}}\cos q}\,dq
     =\frac1{2\pi}\int_0^{2\pi}\!\frac{e^{i|r|q}}{\omega_0-2t_{\mathrm{B}}\cos q}\,dq ,
\end{equation}
where the imaginary part vanishes by symmetry and the evenness $G_r=G_{-r}$ has been used.
We convert this expression to a contour integral using the substitution
\begin{equation}\label{eq:sub}
  \zeta=e^{iq},\qquad dq=\frac{d\zeta}{i\zeta},\qquad
  \cos q=\frac{\zeta+\zeta^{-1}}{2},
\end{equation}
which maps the integral onto the unit circle $|\zeta|=1$ in the complex plane. The denominator
factorizes as
\begin{equation}\label{eq:denomfac}
  \omega_0-2t_{\mathrm{B}}\cos q
  =-\frac{t_{\mathrm{B}}}{\zeta}\Big(\zeta^2-\tfrac{\omega_0}{t_{\mathrm{B}}}\zeta+1\Big)
  =-\frac{t_{\mathrm{B}}}{\zeta}(\zeta-z)(\zeta-z^{-1}),
\end{equation}
whose roots are
\begin{equation}\label{eq:roots}
  z=\frac{\omega_0-\sqrt{\omega_0^2-4t_{\mathrm{B}}^2}}{2t_{\mathrm{B}}},\qquad
  z^{-1}=\frac{\omega_0+\sqrt{\omega_0^2-4t_{\mathrm{B}}^2}}{2t_{\mathrm{B}}},\qquad
  z\,z^{-1}=1 .
\end{equation}
For $0<t_{\mathrm{B}}<\omega_0/2$, both roots are real and positive, with
$0<z<1<z^{-1}$, so only the pole $\zeta=z$ lies inside the unit circle. The contour integral is
therefore
\begin{equation}\label{eq:contour}
  G_r=\frac1{2\pi i}\oint_{|\zeta|=1}
   \frac{\zeta^{|r|}\,d\zeta}{-t_{\mathrm{B}}(\zeta-z)(\zeta-z^{-1})} .
\end{equation}

Applying the residue theorem at the simple pole $\zeta=z$ gives
\begin{equation}\label{eq:residue}
  G_r=\frac{z^{|r|}}{-t_{\mathrm{B}}(z-z^{-1})},\qquad
  z-z^{-1}=-\frac{\sqrt{\omega_0^2-4t_{\mathrm{B}}^2}}{t_{\mathrm{B}}},
\end{equation}
and hence
\begin{equation}\label{eq:Grres}
  G_r=\frac{z^{|r|}}{\sqrt{\omega_0^2-4t_{\mathrm{B}}^2}} .
\end{equation}

Combining \eqref{eq:Vsub} and \eqref{eq:Grres} yields the Born-Oppenheimer potential, Eq.~\eqref{eq:Vr},
\begin{equation}\label{eq:VrBO}
  V(r)
  =-\frac{2\lambda_z^2\,z^{|r|}}{\sqrt{\omega_0^2-4t_{\mathrm{B}}^2}}
  =A\,e^{-|r|/r_0},\qquad
  A=-\frac{2\lambda_z^2}{\sqrt{\omega_0^2-4t_{\mathrm{B}}^2}},\qquad
  r_0=\frac1{\ln(1/z)} .
\end{equation}
At fixed bath parameters, the prefactor scales as $A\propto-\lambda_z^2$, whereas the decay range $r_0$ depends only on the ratio $t_{\mathrm{B}}/\omega_0$ [Eq.~\eqref{eq:Ar0}]. Because hard-core magnons satisfy $r\ge1$, $A=V(0)$ is the extrapolated on-site prefactor rather than an accessible
two-magnon interaction energy. At $r=1$ and in the weak-dispersion limit $t_{\mathrm{B}}\ll\omega_0$,
$z\to t_{\mathrm{B}}/\omega_0$ and
$\sqrt{\omega_0^2-4t_{\mathrm{B}}^2}\to\omega_0$, so
$G_1\to t_{\mathrm{B}}/\omega_0^2$ and
\[
  V(1)\to-2(\lambda_z/\omega_0)^2t_{\mathrm{B}}=-\Vb,
\]
recovering the leading Lang-Firsov term in Eq.~\eqref{eq:Vbind}. Both the range and the magnitude of the
response grow as $t_{\mathrm{B}}\to\omega_0/2$ from below. When
$2t_{\mathrm{B}}=\omega_0$, the bath develops a zero-frequency mode and the displaced ground state is no longer well defined, so Eqs.~\eqref{eq:Grres} and \eqref{eq:VrBO} require the strict stability condition $2t_{\mathrm{B}}<\omega_0$. In the Einstein-bath limit $t_{\mathrm{B}}\to0$, one instead has $z\to0$ and $r_0\to0$; without bath hopping the sites decouple and $V(r)=0$ for $r\ge1$, independently of the on-site interaction $U$.
\section{Hard-core bath limit (\texorpdfstring{$U\to\infty$}{U to infinity})}\label{app:Uinf}

In the limit $U\to\infty$, the weak-source response is analytic: its leading interaction range equals the noninteracting result, while nonlinear response is suppressed. Representing hard-core bosons by Pauli operators,
$\hat a_j\to\hat\sigma^-_j$, $\ad_j\to\hat\sigma^+_j$, and
$\hat n_j\to\tfrac12(1+\hat\sigma^z_j)$. The bath Hamiltonian in Eq.~\eqref{eq:H} then becomes an XX chain in a field,
\begin{equation}\label{eq:HBhc}
  \hat H_{\mathrm{B}}^{\infty}
  =
  -t_{\mathrm{B}}\sum_j
  \left(
    \hat\sigma^+_j\hat\sigma^-_{j+1}+\mathrm{H.c.}
  \right)
  +
  \frac{\omega_0}{2}\sum_j
  \left(
    1+\hat\sigma^z_j
  \right).
\end{equation}
The Jordan-Wigner transformation
$\hat\sigma^-_j=e^{i\pi\sum_{l<j}\hat n_l}\hat f_j$ maps the number-conserving bath Hamiltonian to free
fermions:
\begin{equation}\label{eq:HBff}
  \hat H_{\mathrm{B}}^{\infty}
  =
  \sum_k\varepsilon(k)\,\hat f^\dagger_k\hat f_k,
  \qquad
  \varepsilon(k)=\omega_0-2t_{\mathrm{B}}\cos k .
\end{equation}

The parity-dependent boundary condition of a periodic Jordan-Wigner chain does not affect the thermodynamic limit. Since the one-particle dispersion equals the noninteracting band, the ground state is empty for $\omega_0>2t_{\mathrm{B}}$. At fixed $t_{\mathrm{B}}/\omega_0$, a pinned magnon sources $\lambda_z\hat\sigma^x_m$, and the construction of Appendix~\ref{app:BO} gives
\begin{equation}\label{eq:VrHC}
  V(r)
  =
  -2\lambda_z^2\,\chi_r
  +
  O\!\left(\frac{\lambda_z^4}{\omega_0^3}\right),
  \qquad
  \chi_r
  =
  \bra{0}
  \hat\sigma^x_0
  (\hat H_{\mathrm{B}}^{\infty}-E_0)^{-1}
  \hat\sigma^x_r
  \ket{0},
\end{equation}
where $\chi_r$ is the static transverse response kernel. On the empty bath, $\hat\sigma^x_j\ket{0}=\hat f^\dagger_j\ket{0}$, because the Jordan-Wigner string acts trivially on the
vacuum. The resolvent therefore remains entirely within the one-fermion sector, giving
\begin{equation}\label{eq:chiHC}
  \chi_r
  =
  \frac1L\sum_k\frac{\cos(kr)}{\varepsilon(k)}
  =
  G_r ,
\end{equation}
the same lattice Green's function as at $U=0$. Consequently, to order $\lambda_z^2$, both the depth and the range coincide with the noninteracting result:
\begin{equation}
  V(r)
  =
  -2\lambda_z^2G_r
  +
  O\!\left(\frac{\lambda_z^4}{\omega_0^3}\right)
  \propto
  z^{|r|},
  \qquad
  r_0=\frac{1}{\ln(1/z)}.
\end{equation}
Suppression must therefore arise beyond linear response. With finite transverse sources, the Jordan-Wigner strings become nontrivial outside the one-particle intermediate sector, so the full sourced problem is not quadratic. The nonlinear mechanism is already visible on one site: $\hat a_j+\ad_j$ maps to $\hat\sigma^x_j$, whereas $\hat n_j$ maps to $\tfrac12(1+\hat\sigma^z_j)$. The corresponding harmonic and hard-core Hamiltonians are
\begin{equation}\label{eq:sat_H}
  \hat h_{\mathrm{harm}}
  =
  \omega_0\,\hat n_j
  +
  \lambda_z(\hat a_j+\ad_j),
  \qquad
  \hat h_{\infty}
  =
  \frac{\omega_0}{2}\hat\sigma^z_j
  +
  \lambda_z\hat\sigma^x_j
  +
  \frac{\omega_0}{2}.
\end{equation}
Their ground-state displacements are
\begin{equation}\label{eq:sat_resp}
  \langle\hat a_j+\ad_j\rangle_{\mathrm{harm}}
  =
  -\frac{2\lambda_z}{\omega_0},
  \qquad
  \langle\hat a_j+\ad_j\rangle_{\infty}
  =
  -\frac{2\lambda_z}
  {\sqrt{\omega_0^2+4\lambda_z^2}},
\end{equation}
and their ground-state energies shift by
\begin{equation}\label{eq:sat_E}
  \delta E_{\mathrm{harm}}
  =
  -\frac{\lambda_z^2}{\omega_0},
  \qquad
  \delta E_{\infty}
  =
  \frac{\omega_0}{2}
  -
  \frac12\sqrt{\omega_0^2+4\lambda_z^2}.
\end{equation}
For weak dressing, $\lambda_z\ll\omega_0$, the two responses coincide to leading order, giving
\begin{equation}
  \langle\hat a_j+\ad_j\rangle
  =
  -\frac{2\lambda_z}{\omega_0}
  +
  O\!\left(\frac{\lambda_z^3}{\omega_0^3}\right),
  \qquad
  \delta E
  =
  -\frac{\lambda_z^2}{\omega_0}
  +
  O\!\left(\frac{\lambda_z^4}{\omega_0^3}\right).
\end{equation}
These expansions reproduce the equality of the leading-order induced potentials. For $|\lambda_z|\gtrsim\omega_0$, however, the harmonic response remains linear while the hard-core response saturates:
\begin{equation}
  \langle\hat a_j+\ad_j\rangle_{\infty}
  \longrightarrow
  -\operatorname{sgn}(\lambda_z),
  \qquad
  \delta E_\infty
  =
  -|\lambda_z|
  +
  \frac{\omega_0}{2}
  +
  O\!\left(\frac{\omega_0^2}{|\lambda_z|}\right).
\end{equation}
This reduced response explains the reduced interaction depth. The floor $f_\infty=\lim_{U\to\infty}f(U)$ of Eq.~\eqref{eq:floor} is obtained from the hard-core chain using the Born-Oppenheimer energies with zero, one, and two sources. Equation~\eqref{eq:sat_resp} explains the saturation, but the value of $V(1)$ is not fixed by the single-site response ratio alone: adjacent dressing clouds overlap and contribute jointly.

For Fig.~\ref{fig:VU}, $\lambda_z/\omega_0=0.5$ and $t_{\mathrm{B}}/\omega_0=0.4$ give $f_\infty\simeq0.17$, while $f_\infty\to1$ as $\lambda_z/\omega_0\to0$, consistent with the common linear response. At finite coupling, hard-core saturation does not require $V_\infty(r)/V_{U=0}(r)$ to be exactly independent of $r$. The approximately unchanged fitted range is therefore a numerical property of the regime considered, not an exact consequence of single-site saturation. By contrast, at $t_{\mathrm{B}}\to0$ the flat band gives $G_r=0$ for $r\ge1$, eliminating the intersite interaction.
\section{Numerical methods}\label{app:num}

\subsection{Mean-field integration}

The one- and two-magnon variational equations are integrated on open chains using a fourth-order Runge-Kutta method in double precision. Single-magnon runs start from the normalized Gaussian
\begin{equation}
  \psi_j(0)\propto
  \exp\!\left[-\frac{(j-j_0)^2}{4\sigma^2}\right]e^{ik_0j},
\end{equation}
with $\beta_j(0)=0$ in the polaron frame. This is the natural single-field $D_2$ reference but not the exact transform of a laboratory-frame vacuum for a delocalized magnon; the corresponding bath occupation follows from Eq.~\eqref{eq:nlabdav}. The MPS quench instead starts from the laboratory vacuum. Two-magnon runs use the symmetrized product of separated counterpropagating packets or the compact state
\begin{equation}
  \psi_{jk}(0)\propto
  e^{iK(j+k)/2}
  \exp\!\left[
    -\frac{\big((j+k)/2-X_0\big)^2}{4\sigma^2}
  \right]
  \chi(|j-k|),
\end{equation}
of total momentum $K$, with $\chi(r)$ concentrated near $r=1$. The hard-core condition is imposed through $\psi_{jj}=0$. We record the magnon density $\rho_j$, separation distribution $g(r,t)$, pair weight $P_{\mathrm{nn}}(t)=g(1,t)$, and laboratory-frame cloud $|\gamma_j|^2$ defined in Appendix~\ref{app:dav}. The compact component is tracked by
\begin{equation}
  X_{\mathrm{bound}}(t)
  =
  \frac{
    \sum_j (j+\half)|\psi_{j,j+1}(t)|^2
  }{
    \sum_j |\psi_{j,j+1}(t)|^2
  }.
\end{equation}

The denominator is the instantaneous weight of the nearest-neighbor component, so $X_{\mathrm{bound}}$ follows that component rather than the center of the full two-magnon state. Group velocities are obtained from a linear fit to the packet center over the central half of the propagation window, before boundary reflections become appreciable. For an ideal untruncated Gaussian on the cosine band,
\begin{equation}
  v_{\mathrm{free}}
  =
  v_{\mathrm{bare}}e^{-\sigma_k^2/2},
  \qquad
  v_{\mathrm{bare}}=J\sin k_0,
  \qquad
  \sigma_k=\frac{1}{2\sigma}.
\end{equation}

Finite open chains introduce small discretization, truncation, and boundary corrections. We remove the complete finite-packet offset numerically by comparing each interacting run with a $\lambda_z=0$ control of identical width, central momentum, chain length, and fitting interval. With
\begin{equation}
  v_{\mathrm{LF}}(k_0)
  =
  e^{-\kappa^2}v_{\mathrm{bare}},
\end{equation}
the corrected retardation deficit is
\begin{equation}\label{eq:dv_num}
  \delta v
  =
  1-
  \frac{v}{v_{\mathrm{LF}}(k_0)}
  \frac{v_{\mathrm{bare}}}{v_{\mathrm{free}}}
  =
  1-\frac{v}{e^{-\kappa^2}v_{\mathrm{free}}}.
\end{equation}
For the simulations at $k_0=\pi/2$, $v_{\mathrm{LF}}(k_0)$ coincides with the velocity scale
$v_{\mathrm{LF}}=Je^{-\kappa^2}$ used in the main text. The norm drift and the energy drift,
\begin{equation}
  \epsilon_{\mathcal N}
  =
  \max_t
  \left|
    \sum_j\rho_j(t)-N_{\mathrm{m}}
  \right|,
  \qquad
  \epsilon_E
  =
  \max_t
  \left|
    \mathcal E(t)-\mathcal E(0)
  \right|,
\end{equation}
are monitored, with $N_{\mathrm{m}}=1,2$. Energy conservation tests the Runge-Kutta integrator but not the mean-field factorization, which is assessed against MPS and becomes less reliable as bath-number fluctuations and magnon-boson entanglement grow. The static-XXZ reference fixes $\beta_j=0$, retaining the Lang-Firsov-renormalized hopping and static nearest-neighbor attraction while removing retardation. Stability maps use the same integrator on a fixed coupling grid.

\subsection{Matrix product states}

The MPS representation~\cite{Schollwock2011}, implemented with ITensor and ITensorTDVP~\cite{ITensor2022}, combines each spin and truncated bosonic mode into a composite site,
\begin{equation}
  \mathcal H_j
  =
  \mathbb C^2
  \otimes
  \operatorname{span}
  \left\{
    \ket{0},\ldots,\ket{n_{\max}}
  \right\},
  \qquad
  d_{\mathrm{loc}}=2(n_{\max}+1).
\end{equation}

This keeps the exchange, boson hopping, and longitudinal coupling local or nearest-neighbor in the MPS ordering. Magnetization is used as a $U(1)$ quantum number; boson number is not conserved because the coupling contains $\hat a_j+\ad_j$. The cutoff is selected separately for each parameter set, starting from
\begin{equation}
  n_{\max}^{\mathrm{trial}}
  =
  \min\!\left\{
    10,\,
    \left\lceil
      4+
      8\left(\frac{\lambda_z}{\omega_0}\right)^2
      \left[
        1-\left(\frac{2t_{\mathrm{B}}}{\omega_0}\right)^2
      \right]^{-1/2}
    \right\rceil
  \right\}
\end{equation}
for $2t_{\mathrm{B}}<\omega_0$. It increases with dressing and proximity to the soft-band edge. Production runs use $4\le n_{\max}\le9$; convergence is checked by increasing it and comparing densities, velocities, energies, and entropies. A mean occupation that is small relative to $n_{\max}$ is only an auxiliary diagnostic because it does not guarantee convergence of the Fock-space tail. Ground states and trapped packets are prepared by DMRG in the required fixed-magnetization sector, and compact pairs are launched with
\begin{equation}
  \hat U_K
  =
  \prod_j
  \exp\!\left(
    ik_0j\,\nm_j
  \right),
  \qquad
  K=2k_0,
\end{equation}
applied to the trapped state, with total momentum $K=2k_0$. The collision simulations of Sec.~\ref{sec:bind} use separated counterpropagating Gaussians, and the bath initially occupies its vacuum unless stated otherwise. Two-site TDVP~\cite{Haegeman2011,Haegeman2016,Paeckel2019} uses one two-site sweep per step, SVD cutoff $10^{-9}$, $\Delta t=0.05$-$0.1\,J^{-1}$, and $\chi_{\max}=300$ (single magnon) or $128$ (collisions). Convergence is monitored through energy, total magnetization, maximum link dimension, and the stability of observables under smaller $\Delta t$ and larger $\chi_{\max}$ and $n_{\max}$. Single-magnon runs use $L=64$ and collisions use $L=32$. In the fast-bath limit, the measured single-magnon velocity reproduces the Lang-Firsov narrowing; its residual deficit and $S_{\mathrm{mb}}$ are extracted from the same real-time states, with the results shown in Figs.~\ref{fig:single} and~\ref{fig:retard}. The magnon-boson entropy is
\begin{equation}
  S_{\mathrm{mb}}
  =
  -\Tr
  \left(
    \rho_{\mathrm{s}}\ln\rho_{\mathrm{s}}
  \right),
  \qquad
  \rho_{\mathrm{s}}
  =
  \Tr_{\mathrm{b}}
  \ket{\Psi}\bra{\Psi},
\end{equation}
obtained in the fixed one- or two-magnon basis by contracting the bosonic degrees of freedom and diagonalizing $\rho_{\mathrm{s}}$. It quantifies the full magnon-boson correlations. The single-field Davydov state is a product only in the polaron frame; in the laboratory frame it retains the deterministic adiabatic entanglement of Appendix~\ref{app:collapse}, but not correlations beyond the restricted $D_2$ manifold. By contrast, the spatial bond entropy is
\begin{equation}
  S_{\mathrm{bond}}(\ell)
  =
  -\sum_\alpha
  s_{\alpha,\ell}^2
  \ln s_{\alpha,\ell}^2,
\end{equation}
where $s_{\alpha,\ell}$ are the Schmidt coefficients across the cut $\ell$. Its growth controls the required bond dimension~\cite{HeMillis2017}. It probes a spatial bipartition and is not interchangeable with $S_{\mathrm{mb}}$.

\end{document}